\documentclass[usenatbib,letterpaper]{mn2e}
\usepackage[totalwidth=480pt,totalheight=680pt]{geometry}
\usepackage{psfig}
\usepackage{aas_macros}
\usepackage{amssymb}

\usepackage{color}
\definecolor{dr}{rgb}{0.6,0,0}
\definecolor{db}{rgb}{0,0,0.6}

\usepackage[hyperfootnotes={false},
dvips,
pdftitle={Collisional Evolution of Irregular Satellite Swarms: Detectable Dust around Solar System and Extrasolar Planets.},
pdfauthor={Grant M. Kennedy and Mark C. Wyatt},
pdfstartview={FitH},
bookmarksopen={true} ]{hyperref}
\hypersetup{colorlinks={true}, urlcolor={db}, citecolor={dr}}

\newcommand{\aopt}{$a_{\rm opt}$}

\newcommand{\um}{$\mu$m}
\newcommand{\dc}{$D_{\rm c}$}
\newcommand{\dt}{$D_{\rm t}$}

\newcommand{\dmax}{$D_{\rm max}$}
\newcommand{\dmin}{$D_{\rm min}$}
\newcommand{\mpl}{$M_{\rm pl}$}

\newcommand{\tnl}{$t_{\rm nleft}$}
\newcommand{\mtot}{$M_{\rm tot}$}
\newcommand{\stot}{$\sigma_{\rm tot}$}
\newcommand{\chipr}{$\chi_{\rm PR}$}
\newcommand{\qd}{$Q_{\rm D}^\star$}
\newcommand{\fq}{$f_{\rm Q}$}
\newcommand{\fv}{$f_{\rm v_{\rm rel}}$}
\newcommand{\vrel}{$v_{\rm rel}$}
\newcommand{\rj}{$R_{\rm \jupiter}$}
\newcommand{\mj}{$M_{\rm \jupiter}$}
\newcommand{\rh}{$R_{\rm H}$}
\newcommand{\apl}{$a_{\rm pl}$}
\newcommand{\ndt}{$n(D_{\rm c}/2 \rightarrow D_{\rm c})$}
\newcommand{\nstr}{$N_{\rm str}$}
\newcommand{\jupiter}{Jup}

\begin{document}

\title[Evolution of Irregular Satellite Swarms]{Collisional Evolution of Irregular
  Satellite Swarms: Detectable Dust around Solar System and Extrasolar Planets.}

\author[G. M. Kennedy \& M. C. Wyatt]
{G. M. Kennedy\thanks{Email: \href{mailto:gkennedy@ast.cam.ac.uk}{gkennedy@ast.cam.ac.uk}} and M. C. Wyatt \\
  Institute of Astronomy, University of Cambridge, Madingley Road,
  Cambridge CB3 0HA, UK}

\maketitle

\begin{abstract}
  Since the 1980's it has been becoming increasingly clear that the Solar System's
  irregular satellites are collisionally evolved. The current populations are remnants of
  much more massive swarms that have been grinding away for billions of years. Here, we
  derive a general model for the collisional evolution of an irregular satellite swarm
  and apply it to the Solar System and extrasolar planets. The model uses a particle in a
  box formalism and considers implications for the size distribution, which allows a
  connection between irregular satellite populations and predicted levels in the
  resulting dust cloud. Our model reproduces the Solar System's complement of observed
  irregulars well, and suggests that the competition between grain-grain collisions and
  Poynting-Robertson (PR) drag helps set the fate of the dust.  In collision dominated
  swarms most dust is lost to interplanetary space or impacts the host planet, while PR
  dominated grains spiral in towards the planet through the domain of regular
  satellites. Because swarm collision rates decrease over time the main dust sink can
  change with time, and may help unravel the accretion history of synchronously rotating
  regular satellites that show brightness asymmetries, such as Callisto and Iapetus. Some
  level of dust must be present on AU scales around the Solar System's giant planets if
  the irregular satellites are still grinding down, which we predict may be at detectable
  levels. We also use our model to predict whether dust produced by extrasolar
  circumplanetary swarms can be detected. Though designed with planets in mind, the
  coronagraphic instruments on JWST will have the ability to detect the dust generated by
  these swarms, which are most detectable around planets that orbit at many tens of AU
  from the youngest stars. Because the collisional decay of swarms is relatively
  insensitive to planet mass, swarms can be much brighter than their host planets and
  allow discovery of Neptune-mass planets that would otherwise remain invisible. This
  dust could have been detected by HST ACS coronagraphic observations, and in one case
  dust may have already been detected. The observations of the planet Fomalhaut b can be
  explained as scattered light from dust produced by the collisional decay of an
  irregular satellite swarm around a $\sim$10\,$M_\oplus$ planet. Such a swarm comprises
  about 5 Lunar masses worth of irregular satellites. Finally, we briefly consider what
  happens if Fomalhaut b passes through Fomalhaut's main debris ring on a coplanar orbit,
  which allows the circumplanetary swarm to be replenished through collisions with ring
  planetesimals. This scenario, in which the planet is at least of order an Earth mass,
  may be ruled out by the narrow structure of the debris ring.
\end{abstract}

\begin{keywords}
  circumstellar matter --  stars: planetary systems: general -- Solar System: planets and satellites
\end{keywords}

\section{Introduction}\label{s:intro}

With a penchant for retrograde, barely-bound, high-eccentricity orbits and
flatter-than-usual size distributions, the irregular satellites are one of the Solar
System's rebel populations. Their presumed capture into these unusual orbits around the
Solar System's giant planets has long been a puzzle. Because irregulars exist at all four
outer planets, capture mechanisms specific to formation of gas giants like Jupiter and
Saturn \citep[e.g.][]{1977Icar...30..385H,1979Icar...37..587P} are not general
enough. Dynamical mechanisms, which do not rely on the existence or growth of large
gaseous atmospheres, can be applied to gas and ice-giants alike and are therefore
favoured \citep[e.g.][]{1971Icar...15..186C}. \citet{2007AJ....133.1962N} recently
proposed a different dynamical mechanism as part of a unified model of outer Solar System
formation \citep{2005Natur.435..466G,2005Natur.435..459T}, where irregulars are captured
during a period of instability via planet-planet interactions.

An apparent weakness of the planet-planet interaction mechanism is that the irregular
satellites should share the same size distribution as Jupiter's Trojan asteroids, which
in this model were captured at the same time from the same population
\citep{2005Natur.435..462M}. In fact the differences in the Trojan and irregular
satellite size distributions are marked, with irregular satellites being much flatter for
sizes larger than about 10\,km. To overcome this hurdle, \citet{2010AJ....139..994B}
showed that while the size distribution was indeed \emph{initially} steeper like the
Trojans, 4.5\,Gyr or so of collisional evolution is sufficient to reduce primordial
irregular satellite populations to a size distribution that matches those currently
observed. This result implies that there were previously more irregular satellites and
perhaps most significantly, that copious amounts of dust were produced during the
depletion of these satellites.

The evidence for collisional evolution of irregular satellites has been mounting for some
time. \citet{1981Icar...48...39K} showed that the four prograde irregulars known to orbit
Jupiter at the time had a relatively short collisional lifetime. The advent of
large-format CCD surveys since the turn of the century has seen a dramatic increase the
number of irregulars and made further theoretical advances possible
\citep[e.g.][]{2001Natur.412..163G,2003Natur.423..261S,2004Natur.430..865H,2007ARA&A..45..261J}. \citet{2003AJ....126..398N}
noted that irregular satellites around planets closer to the Sun have larger orbits (in
Hill radii), suggesting that the lack of satellites closer to Jupiter is due to their
erosion through collisions, which proceed at a faster rate closer to the planet. Using
numerical integrations to derive average orbital elements, they also proved the existence
of collisional families \citep[see also ][]{2001Natur.412..163G}. This latter discovery
is particularly important, because the current collision rate amongst the irregular
satellites is in some cases too low to explain their existence
\citep{2003AJ....126..398N}. The inference is again that the number of irregular
satellites, and thus their collision rate, was much greater in the past and that
they decayed through collisions to the current level.

Coinciding with these theoretical advances was the discovery by direct imaging of
Fomalhaut b \citep{2008Sci...322.1345K}, an extrasolar planet predicted to exist based on
the elliptical orbit of Fomalhaut's circumstellar debris ring
\citep{2005Natur.435.1067K,2006MNRAS.372L..14Q}. The ring structure suggests that the
planet is less than 3\,\mj, though the planet could be much less massive
\citep{2009ApJ...693..734C}. This discovery appears unrelated to irregular satellites,
but the inability of Hubble, Keck, and Gemini photometry to pin down whether the planet
looks like a planetary atmosphere or reflected starlight provides the link.

While the planets discovered to orbit HR 8799 appear to be consistent with $\sim$1000K
substellar-mass objects \citep{2008Sci...322.1348M}, multi-wavelength photometry of
Fomalhaut b appears bluer than expected for a 200\,Myr old gas-giant
planet. Specifically, Fomalhaut b has so far defied detection at wavelengths longer than
1\um, leading \citet{2008Sci...322.1345K} to suggest that the spectrum is actually
starlight scattered from an optically thick circumplanetary disk of about 20 Jupiter
radii. Though such a scenario is plausible, we argue that dust produced by a swarm of
colliding irregular satellites is also a possibility.

Given that the Solar System's irregular satellite complement decayed to its current state
through collisions and the exciting possibility that Fomalhaut b may harbour the first
circumplanetary dust seen outside the Solar System, the time seems right to consider
whether such clouds of irregular satellites could be visible around extrasolar
planets. In the following sections, we derive a simple model for the evolution of a
circumplanetary satellite swarm and the all-important extrasolar observable---the
dust. We compare our model with the Solar System irregulars, and comment on the fate and
observability of dust. We then apply the model to circumplanetary swarms around
extrasolar planets. Finally, we explore what kind of satellite swarm could exist around
Fomalhaut b, and what constraints this proposed swarm puts on the planet mass.

\section{Model of a Circumplanetary Swarm}\label{s:mod}

The irregular satellite swarms described in this paper have not knowingly been detected
around other planets that orbit other stars. Therefore, like the pre-1995 days of planet
formation theory, we must take cues from the Solar System. However, based on experience
gained from the surprising diversity of extrasolar planets, we should not assume that our
irregular satellite complement is typical, or that extrasolar analogues should follow all
the same rules.

Thankfully, some of the most important irregular satellite properties are dynamical and
would have been discovered even if the Solar System had no irregulars. The main dynamical
curiosity is their inclinations, which are all within about 60$^\circ$ of the ecliptic
(but include retrograde orbits). This evacuation of near-polar orbits is due to Solar and
planetary perturbations, which drive the eccentricities of highly inclined orbits to such
large values that they either encounter regular satellites or leave the Hill sphere
\citep{2002Icar..158..434C,2003AJ....126..398N}.

Another constraint comes from the stability of circumplanetary orbits. Although
\citet{2008AJ....136.2453S} find that satellites out to a few Hill radii could survive
the age of the Solar System around Uranus and Neptune, all currently known irregulars
have orbits with semi-major axes $a_{\rm pl}$ less than half the Hill radius \rh\ \citep[
e.g.][]{2003Natur.423..261S,2004Natur.430..865H,2005AJ....129..518S}
\begin{equation}\label{eq:rh}
  R_{\rm H} = a_{\rm pl} \left( M_{\rm pl} / 3\,M_\star \right)^{1/3} \, ,
\end{equation}
where $M_{\rm pl}$ is the planet mass and $M_\star$ the stellar mass. On the sky,
  the giant planets' Hill radii span several degrees. The exact stability limit has a
small inclination dependence in that retrograde orbits are stable at larger distances
than the widest stable prograde orbits \citep{2003AJ....126..398N}.

In this section we outline a model for the collisional evolution of irregular satellite
swarms. Because we want to make predictions of the only possible extrasolar
observable---dust---we keep our model simple. There are many uncertainties in
extrapolating a swarm of irregular satellites to a cloud of dust, such as the strength of
satellites, the size distribution slope, and the minimum grain size, which at this stage
make the development of a more complex collisional model largely unnecessary.

The next four subsections contain many equations that describe properties of a
circumplanetary swarm. Readers looking for actual numbers may like to refer ahead to
Table \ref{t:ssdust}, which shows estimates of some properties for the Solar System giant
planets.

\subsection{Collisional mass loss}\label{ss:decay}

In a steady-state collisional cascade the mass within a given size range decreases as
these objects are destroyed in collisions, but is replaced at the same rate by fragments
created by destruction of larger objects. Mass is lost at the bottom end of cascade,
usually by radiation forces that remove grains smaller than some minimum size. The
evolution of the size distribution is therefore dictated by the collisional decay of the
largest objects.

The result of such a collisional cascade would, in an ideal situation (cascade infinite
in extent, strength independent of size), have a steady state size distribution with a
well defined slope of $n(D) = K \, D^{2-3q}$ where $n(D)dD$ is the number of satellites
between $D$ and $D+dD$ and $q=11/6$ \citep{1969JGR....74.2531D}.

In fact the true distribution of circumstellar collisional cascades like the asteroid
belt is expected to have different slopes in different size ranges due to the way
strength depends on size \citep{2003Icar..164..334O}. Strength is typically described by
the parameter \qd, which is the kinetic collision energy per target mass needed to
shatter \emph{and disperse} the target, such that the largest remnant is half the mass of
the original target. Such a collision is commonly termed ``catastrophic.''

Small objects are held together by their own material strength, and grow weaker with
increasing size due to the increased likelihood of the presence of a significant flaw. To
quote \citet{1994Icar..107...98B}, ``Subdivide this same rock into 100 equal pieces and
99 of them are now stronger than the original, owing to the simple fact that they do not
contain the one weakest flaw.'' Above some transition size ($D_{\rm t} \sim 0.1km$)
bodies gain strength from self-gravity. The energy needed for catastrophic disruption now
increases with size. Though objects may be shattered, extra energy is required to ensure
the fragments have sufficient escape speeds and are no longer bound. Gravity also limits
fracture propagation within the material, thus adding strength
\citep[e.g.][]{1999Icar..142....5B}. This behaviour is usually modelled using complex
numerical codes, and parameterised by a power law for each of the strength and gravity
regimes. In fact \qd\ varies by about a factor 10 over the range of impact parameters and
there are similar differences between strength laws derived by different studies
\citep[e.g.][]{1999Icar..142....5B,2009ApJ...691L.133S}. Thus \qd\ and the resulting size
distribution are the most uncertain inputs for our model.

We set planetesimal strength with the \citet{1999Icar..142....5B} law for ice at 3\,km
s$^{-1}$. For objects larger than $D_{\rm t} = 0.1$ (in km),
\begin{equation}\label{eq:qd}
  Q_{\rm D}^\star = 0.1\, \rho \, D^{1.26} / f_{\rm Q}
\end{equation}
in J kg$^{-1}$ where the mass density $\rho$ is in kg m$^{-3}$. The strength dependence
of small objects ($\propto$$D^{-0.39}$) is only used in setting the size distribution of
objects smaller than \dt. Following \citet{2010AJ....139..994B}, we allow objects to be
weaker than the \citet{1999Icar..142....5B} law by including the factor \fq\ \citep[see
also][]{2009Natur.460..364L}. The strength law is similar to the
\citet{2009ApJ...691L.133S} strength law when $f_{\rm Q} =
8$. \citet{2010AJ....139..994B} reproduce the Solar System's irregular satellite
populations best when $f_{\rm Q} > 3$ so we set $f_{\rm Q} = 5$.

The size distribution of objects with such strength properties is expected to have a
slope with $q_{\rm{s}}=1.9$ at $D<D_{\rm t}$ (in the strength-dominated regime) and
$q_{\rm{g}}=1.7$ for $D>D_{\rm t}$ \citep[in the gravity-dominated
regime,][]{2003Icar..164..334O}. Although several wiggles are also expected in the
distribution \citep{1994P&SS...42.1079C,1998Icar..135..431D}, here the size distribution
is assumed to be continuous with the appropriate slopes ($q_{\rm{s}}$ or $q_{\rm{g}}$)
between the smallest objects of size $D_{\rm{min}}$ (in \um) and the largest objects
participating in the collisional cascade of size $D_{\rm{c}}$ (in km). As we show later,
\dc\ may be smaller than the largest object, which has size \dmax.

With this two phase size distribution, the conversion between the size distribution's
surface area (\stot\ in AU$^2$) and mass (\mtot\ in $M_\oplus$) is
\begin{equation}\label{eq:mtotgen}
  M_{\rm tot} = 0.0025 \, \rho \, \sigma_{\rm tot} \, \frac{3q_{\rm s}-5}{6-3q_{\rm g}} \,
  D_{\rm c}^{6-3q_{\rm g}} \, D_{\rm t}^{3q_{\rm g}-3q_{\rm s}} \, D_{\rm min}^{3q_{\rm s}-5}
\end{equation}
where we assume $D_{\rm min} \ll D_{\rm t} \ll D_{\rm c}$. For $q_{\rm s} = 1.9$, $q_{\rm
  g} = 1.7$, and $D_{\rm t} = 0.1$ (in km), this relation simplifies to
\begin{equation}\label{eq:mtot}
  M_{\rm tot} = 3.9 \times 10^{-6}\, \rho\, \sigma_{\rm tot} \,D_{\rm c}^{0.9}\, D_{\rm
    min}^{0.7} \, .
\end{equation}
This mass only includes objects between \dmin\ and \dc\ (not \dmax) because we use \mtot\
below to calculate collision rates.

The collisional lifetime of satellites of size $D_{\rm{c}}$ can be calculated from the
total mass using the particle-in-a-box approach. Here we follow
\citet{2010MNRAS.402..657W}, who also take into account that objects of size \dc\ can be
destroyed in catastrophic collisions by those down to a size $X_{\rm{c}}\,D_{\rm c}$
(assumed to be $\gg 0.1$km) where $X_{\rm c} = \left( 2 Q_{\rm D}^\star / v_{\rm rel}^2
\right)^{1/3}$ and the collision velocity \vrel\ is in m s$^{-1}$. The rate of
catastrophic collisions is
\begin{equation}\label{eq:rcc}
  R_{cc} = 8.4 \times 10^{-5} \, \frac{6-3q_{\rm g}}{3q_{\rm g}-5}\,\frac{ v_{\rm{rel}} \, C_1
    X_{\rm c}^{C_2}\, M_{\rm tot} }
  { \rho \, D_{\rm c} \, V }
\end{equation}
in years$^{-1}$, where $V$ is the volume occupied by the satellites in AU$^3$. The
inverse of the collision rate is called the collision time $t_{\rm c}$. The assumption of
$D_{\rm t} \ll X_{\rm{c}}\,D_{\rm c}$ means that only $q_{\rm g}$ is needed for the
collision rate. To obtain equation (\ref{eq:rcc}), it was necessary to integrate over the
size distribution from $X_{\rm c}\,D_{\rm c}$ to \dc, which yields a function called
$G(q,X_{\rm c})$ by \citet{2007ApJ...658..569W}. In \citet{2010MNRAS.402..657W} the
approximation $G \left( 11/6,X_c \right) \approx C_1 X_c^{C_2}$ with $C_1 = 0.2$ and $C_2
= -2.5$ is used (the limit of small $X_c$). Substituting these values and $q_{\rm g} =
11/6$ yields Equations (9) \& (10) of \citet{2010MNRAS.402..657W}.  Here, we find a
numerical approximation for the function $G \left( q,X_c \right)$, which is within 10\%
for the more physically plausible range $0.01 < X_c < 0.75$ and $1.7 < q <2$, yielding
$C_1 = 2.62(q-1.66)$ and $C_2 = 2.70(0.98-q)$. For $q_{\rm g} = 1.7$, $C_1 = 0.1$ and
$C_2 = -1.9$.

Irregular satellites are assumed to orbit the planet at semi-major axes relative to the
Hill radius in the range $\eta \pm d\eta/2$ (we use $d\eta = \eta/2$). The volume the
satellites occupy is $V = 4 \pi \eta_{\rm{s}}^2 d\eta_{\rm{s}} r_{\rm{H}}^3 \times
0.866$, where the extra factor of $0.866$ accounts for the lack of near polar orbits.

The mean relative velocity of collisions is expected to be some fraction \fv\ of the
Keplerian velocity at $\eta$, which (in m s$^{-1}$) is
\begin{equation}\label{eq:vk}
  v_{\rm k} = 516 \, M_{\rm pl}^{1/3} \, M_\star^{1/6} / ( \eta \, a_{\rm pl} )^{1/2} \, .
\end{equation}
That fraction will depend on the eccentricities and inclinations of the satellite swarm.
A simple estimate of the mean collision velocity comes from assuming circular and
isotropic orbits, yielding $f_{\rm v_{\rm rel}} = 4/\pi$ and typical impact velocities of
$\sim$0.5--3\,km/s for Solar System giant planets. These velocities are high enough that
the impactor/target mass ratio for catastrophic collisions is small, so the energy lost
in a collision is also small. Thus very little kinetic energy is lost in a typical
collision and collisional damping is unimportant. Using the Monte-Carlo eccentric ring
model of \citet{2010MNRAS.402..657W}, we find that the mean collision velocity is similar
for a realistic orbital distribution, and somewhat lower when eccentricities are
introduced. The values vary between about 0.9--1.3, so we adopt $f_{\rm v_{\rm rel}} =
4/\pi$ throughout. In fact orbits have a range of inclinations and eccentricities, and
each collision has a different probability, which is itself a function of the relative
velocity \citep{1994Icar..107..255B}.

Substituting our approximation for $C_1$ and $C_2$ for $q_{\rm g} = 1.7$ yields the rate
of catastrophic collisions
\begin{equation}\label{eq:rccdc}
  R_{cc} = 1.3 \times 10^7 \,
  \frac{ M_{\rm tot} \, M_\star^{1.38} \, f_{\rm v_{\rm rel}}^{2.27} }{ \, {Q_{\rm D}^\star}^{0.63}
    \rho \, D_{\rm c} \, M_{\rm pl}^{0.24} \, (a_{\rm pl} \, \eta)^{4.13} }
\end{equation}
in years$^{-1}$. As one expects, the rate is independent of $a_{\rm pl}$ for the same
physical swarm (i.e doubling $a_{\rm pl}$ halves $\eta$). The factors that largely set
the collision rate are $\eta$, \dc, $a_{\rm pl}$, and \fv. The strongest contributions
are from $\eta$ and $a_{\rm pl}$, which set the cloud volume and space density. The rate
also depends strongly on the mean collision velocity because this speed sets both the
rate at which an object sweeps through space and the number of impactors that result in a
catastrophic disruption. Greater collision velocities mean smaller impactors can destroy
a given object, and smaller impactors are more numerous. Because \qd\ also depends on
\dc\ (eq. \ref{eq:qd}), the collision rate also depends strongly on \dc. For the same
total mass, larger \dc\ means fewer large objects, which are also stronger.

Curiously, the least important parameter is the mass of the planet itself. This result
arises because in equation (\ref{eq:rccdc}), \vrel\ is slightly less than linearly
dependent on \mpl, which nearly cancels with the linear dependence of volume on \mpl. In
simple terms, for fixed $\eta$ and \mtot\ the increase in volume with planet mass works
out to be slightly stronger than the increase in velocity, so the collision rate
decreases slowly as the planet mass increases.

We validate our model with the eccentric ring model \citep{2010MNRAS.402..657W}. Our
collision rate is within a factor of three for a range of eccentricities with the best
agreement for high values, sufficient for our purposes here considering that much larger
uncertainty lies with assumed material properties and the resulting size distribution.

\subsection{Radiation forces on dust}\label{ss:rad}

To derive the surface area in small grains (dominated by small objects) from the total
mass (dominated by large objects) we need to know the size of the smallest grains that
can survive in circumplanetary orbits. Two notable detections are micron-sized grains
found orbiting at large (50--350\,\rj) distances from Jupiter
\citep{2002Icar..157..436K,2010P&SS...58..965K}, and the large tenuous ring found
orbiting Saturn \citep{2009Natur.461.1098V}. Both studies attribute material released by
impacts from interplanetary grains as a likely source, though
\citeauthor{2009Natur.461.1098V} note that debris from irregular satellite collisions
impacting Phoebe could also be the cause.

As with grains orbiting a star, the effect of radiation forces on dust characterised by
$\beta = F_{\rm radiation}/F_{\rm gravity}$ (both due to the star) plays the most
important role in setting the minimum size of grains that survive on circumplanetary
orbits \citep{1979Icar...40....1B}. Other effects related to interaction with planetary
magnetospheres \citep[e.g.][]{1996ARA&A..34..383H} play some part but are less important
for the $\gtrsim$\um\ grain sizes and wide orbits considered here.

\subsubsection{Radiation pressure}

Radiation pressure is the radial component of the force, which in contrast to
circumstellar orbits causes the orbits of dust grains to evolve. While semi-major axes
remain constant, eccentricities oscillate with a period equal to the planet's orbital
period, with a maximum that depends on $\beta$ and the grain orbit. The maximum $\beta$
of grains that survive in orbit around the planet with $e < 1$ have
\citep{1979Icar...40....1B}
\begin{equation}\label{eq:betac}
  \beta_{\rm c} = v / 3 v_\odot =
  5.8 \times 10^{-3} M_{\rm pl}^{1/3} / ( M_\star^{1/3} \eta^{1/2} ) \, ,
\end{equation}
where $v$ is the velocity of a grain as it orbits the planet, and $v_\odot$ is the
velocity of the planet as it orbits the star. For typical planets and irregular satellite
orbits $\beta_{\rm c}$ is much smaller than the blowout limit of 0.5 for stellocentric
orbits.

Because $\beta$ for normal grains peaks where the star radiates most of its radiation, it
might be possible for $\beta_{\rm c}$ to allow both large and very small grains to
survive, with only grains in the peak being excluded. However, for sub-micron
``astronomical silicate'' grains, those on the small side of the peak, the smallest
grains have $\beta \sim 0.11$ \citep[e.g.][]{1994AREPS..22..553G} and is higher for more
massive stars, so grains smaller than the wavelength of typical stellar radiation will
usually be ejected.

To convert $\beta_{\rm c}$ into a minimum size, we use $D_{\rm min} = (1150/\rho
\beta_{\rm c})L_\star/M_\star$ in $\mu$m \citep[e.g.][]{2008arXiv0807.1272W}. The minimum
size is therefore
\begin{equation}\label{eq:dmin}
  D_{\rm min} = 2 \times 10^5 \frac{ \eta^{1/2} \, L_\star }{ \rho \, M_{\rm pl}^{1/3} \, M_\star^{2/3}}
\end{equation}
in \um. For typical parameters, \dmin\ is at least \um-size. For Jupiter and Neptune,
Equation (\ref{eq:dmin}) yields 12 and 23\um, an order of magnitude larger than the
minimum (blowout) size for the same grains on circumsolar orbits.

\subsubsection{Poynting-Robertson drag}

An alternative to grain removal by radiation pressure is orbital decay due to
Poynting-Robertson (PR) drag. The decay timescale is similar to the heliocentric case
\begin{equation}\label{eq:tpr}
  t_{\rm PR} = 530 \, a_{\rm pl}^2 / ( \beta \, M_\star )
\end{equation}
in years \citep{1979Icar...40....1B}. However, as noted above $\beta$ for the smallest
grains is typically much smaller than 0.5 and $t_{\rm PR}$ correspondingly longer.

For grains to spiral into the planet by PR drag they must avoid colliding with other
grains first, which breaks them into smaller particles that are instead removed by
radiation pressure. The competition between PR drag and collisions can be characterised
by $\chi_{\rm PR} = t_{\rm PR} / t_{\rm col}$. \citet{1999ApJ...527..918W} showed that
for the smallest grains this collision rate is roughly $t_{\rm per} r_{\rm dust}^2 / (4\,
\sigma_{\rm tot})$ for a flat disk with radial extent $r_{\rm dust} \pm r_{\rm dust}/4$
(where $t_{\rm per}$ is orbital period). Adapting this expression to an isotropic case
results in a small change due to the greater cloud volume and faster collision
velocities: $t_{\rm col,dust} = t_{\rm per} r_{\rm dust}^2 / (4 \, f_{v_{\rm rel}}
\sigma_{\rm tot})$, or using our parameters
\begin{equation}\label{eq:tcoldust}
  t_{\rm col,dust} = 10^{-5} \, \frac{ \left( \eta \, a_{\rm pl} \right)^{7/2} M_{\rm
      pl}^{2/3} } { M_\star^{7/6} \, \sigma_{\rm tot} }
\end{equation}
in years. Expressed in terms of our basic parameters, the ratio is
\begin{equation}\label{eq:chipr}
  \chi_{\rm PR} = 4 \times 10^{4} \, \frac{ \rho \, D_{\rm min} \, \sigma_{\rm tot} \, M_\star^{7/6} }
  { a_{\rm pl}^{3/2} \, \eta^{7/2} \, M_{\rm pl}^{2/3} \, L_\star }
\end{equation}
When this ratio is larger than unity, grains suffer collisions before their orbits have
time to decay due to PR drag, and are subsequently removed by radiation pressure (termed
``collision dominated''). When this ratio drops below unity, grains spiral in towards the
planet before they collide and may encounter any existing regular satellites as they do
so (``PR dominated''). Substituting Equation (\ref{eq:dmin}) for \dmin\ yields
\begin{equation}\label{eq:chipr2}
  \chi_{\rm PR} = 8 \times 10^9 \, \frac{ \sigma_{\rm tot} \, M_\star^{1/2} }
  { a_{\rm pl}^{3/2} \, \eta^3 \, M_{\rm pl} } \, .
\end{equation}
With only the Solar System example to go by (see \S \ref{ss:ssdust}), it is impossible to
tell whether the ``typical'' satellite dust cloud will be collision or PR
dominated. However, due to the amount of dust needed for an extrasolar irregular
satellite swarm to be detectable ($\sigma_{\rm tot} \gtrsim 10^{-4}$\,AU$^2$), any
observed extrasolar swarms will likely be collision dominated.

\subsubsection{Summary}

In order to estimate the minimum grain size we have necessarily made a number of
simplifications. We have used expressions for radiation forces assuming low inclination
orbits akin to Saturn's ring. In fact $\beta_{\rm c}$ varies with the orientation of the
orbit relative to the Solar direction and is smallest for those with pericenter initially
aligned with the Solar direction. Equation (\ref{eq:betac}) assumes a coplanar orbit with
pericenter perpendicular to the Solar direction and underestimates $\beta_{\rm c}$ for
certain loss by a factor 2--3 \citep{1979Icar...40....1B}. We therefore overestimate the
minimum grain size, and underestimate \stot.

Our prescription for \dmin\ does not preclude detection of grains smaller than
\dmin. Grains below the minimum size may complete many orbits before their eccentricity
exceeds unity. Also, the minimum grain size decreases with $\eta$ so small bound
eccentric grains with smaller planetocentric semi-major axes can be found anywhere within
$2\eta$. On a detailed level, \dmin\ takes on a range of values and for the smallest
grains with $\beta \sim \beta_{\rm c}$, the orbit averaging used in deriving Equation
(\ref{eq:betac}) breaks down. Using numerical simulations, \citet{2002Icar..157..436K}
find that \dmin\ for Jupiter is $\sim$1\,\um\ (compared to our value of 12.4\,\um, see
Table \ref{t:ssdust}). Considering that our value is overestimated by the factor 2-3
noted above, the minimum size is probably 1 to a few \um, with differences in the assumed
grain properties contributing some uncertainty.

While our simplifications are reasonable, they gloss over important aspects of grain
dynamics. For example, \citet{2002Icar..157..436K} show circumplanetary dust clouds to
have both size-dependent and pro/retrograde orbit sensitive structure. Because
non-gravitational forces cause small objects to deviate from the orbits of their parent
bodies, such forces lead to effects beyond the scope of our model. Therefore, future work
will need to consider how grain orbital evolution affects both the physical appearance of
the cloud, and the underlying size distribution. For example, in a typical collision
dominated circumstellar disk the minimum (blowout) grain size is a single number,
independent of radial distance. This fairly sharp truncation creates a wave in the size
distribution due to the alternating lack and then over-abundance of projectiles that
destroy larger targets \citep{1994P&SS...42.1079C}. However, in a circumplanetary cloud
the minimum grain size varies with circumplanetary distance (Eq. \ref{eq:dmin}), which
could lead to qualitatively different and spatially varying size distributions. In \S
\ref{ss:ssdisc} we suggest that Jupiter's swarm may be PR dominated, which can lead to
further differences at the small end of the size distribution. In addition, other
non-gravitational forces known to be important for circumstellar dynamics
\citep[e.g. Yarkovsky effect,][]{1979Icar...40....1B} may be important for the evolution
of circumplanetary orbits and consequent collisional evolution.

\subsection{Time evolution}\label{ss:evol}

We now turn to the time evolution of disk properties. Assuming that the size distribution
remains fixed and that mass is lost by catastrophic disruption of the largest objects,
the disk mass remaining as a function of time is found by solving
\citep[e.g.][]{2007ApJ...658..569W}
\begin{equation}\label{eq:dmdt}
  \frac{dM_{\rm tot}}{dt} = -M_{\rm tot} \, R_{\rm cc} \, ,
\end{equation}
which yields
\begin{equation}\label{eq:mtime}
  M_{\rm tot}(t) = \frac{ M_{\rm tot}(0) }{ 1 + R_{\rm cc}(0) \, t }
\end{equation}
where we use $t$ or 0 to indicate variables at a particular time where necessary. The
initial cloud mass is $M_{\rm tot}(0)$ and $M_{\rm tot}(t) = M_{\rm tot}(0) / R_{\rm
  cc}(0) \, t$ when $R_{\rm cc}(0)\,t \gg 1$. Equation (\ref{eq:mtot}) shows that mass
and surface area are linearly proportional for fixed \dc\ and \dmin\ so the surface area
of small grains decays in the same way. Because the collision rate depends on the
remaining mass, after a few collision times the remaining mass is independent of the
initial mass \citep{2007ApJ...658..569W}.

\subsubsection{Where are the most massive swarms?}\label{ss:aopt}

There is an interesting interplay between the initial cloud mass and planet semi-major
axis. For arbitrarily large initial masses, swarms at any distance and time are on the
$1/t$ part of their evolution where the remaining mass is independent of the initial mass
\citep[called a ``collision limited'' disk by][]{2010MNRAS.401..867H}. Because these
systems have a maximum remaining mass that depends on the collision time ($1/R_{\rm cc}$)
without the mass term, the remaining mass increases strongly with planet semi-major
axis. In fact, $t_{\rm c} \propto a_{\rm pl}^{4.13}$ (i.e. more strongly than $a_{\rm
  pl}^2$), so collision limited swarms are brighter in scattered light for larger planet
semi-major axes (for fixed planet mass, age, etc.).

Of course, the initial swarm mass cannot be arbitrarily large, so swarms around
sufficiently distant planets will take some time to start colliding. Thus, for fixed
initial mass, swarms around close planets will rapidly decay due to the short collision
times, while more distant planets all have the same mass in satellites because the
largest irregulars have not yet, or only just begun to collide. Of these more distant
planets, the one whose swarm has just started to suffer collisions is the one that
receives the most stellar insolation and is brightest in scattered light.

The semi-major axis of the planet hosting this swarm can easily be worked out from the
collision rate in equation (\ref{eq:rccdc}), because a swarm that has just started to
collide has $t = 1/R_{\rm cc}$, therefore
\begin{equation}\label{eq:aopt}
  a_{\rm opt} = 50 \, \frac{ M_\star^{0.33} \, f_{v_{\rm rel}}^{0.55} }
  { M_{\rm pl}^{0.06} \, {Q_{\rm D}^\star}^{0.15} \, \eta }
  \left( \frac{  t \, M_{\rm tot}(0) }{ \rho \, D_{\rm c} } \right)^{0.24}
\end{equation}
in AU. The ``opt'' subscript indicates that this planet lies at the optimum distance to
be detected at ``opt''-ical wavelengths. This prescription for the brightest swarm is
complicated by the fact that faint objects are harder to detect close to host stars, an
issue we return to in \S \ref{ss:exo}. The same concept of an optimal distance applies to
thermal emission, but is somewhat different because the cloud temperature changes with
planet semi-major axis.

\subsubsection{Stranding the largest objects}\label{ss:stranding}

\begin{figure*}
  \begin{center}
    \vspace{-0.2in}
    \hspace{-0.35in} \psfig{figure=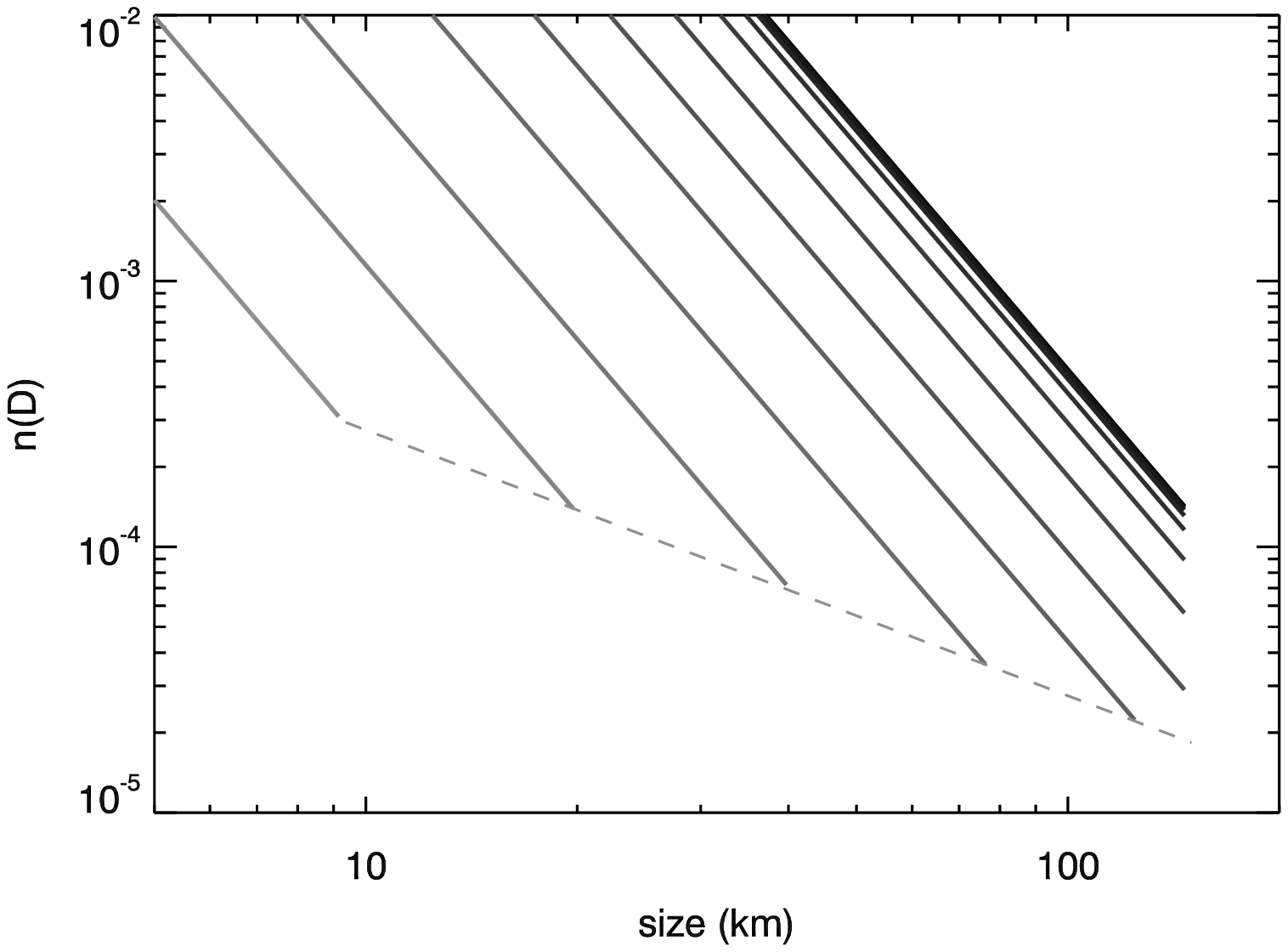,width=0.52\textwidth}
    \psfig{figure=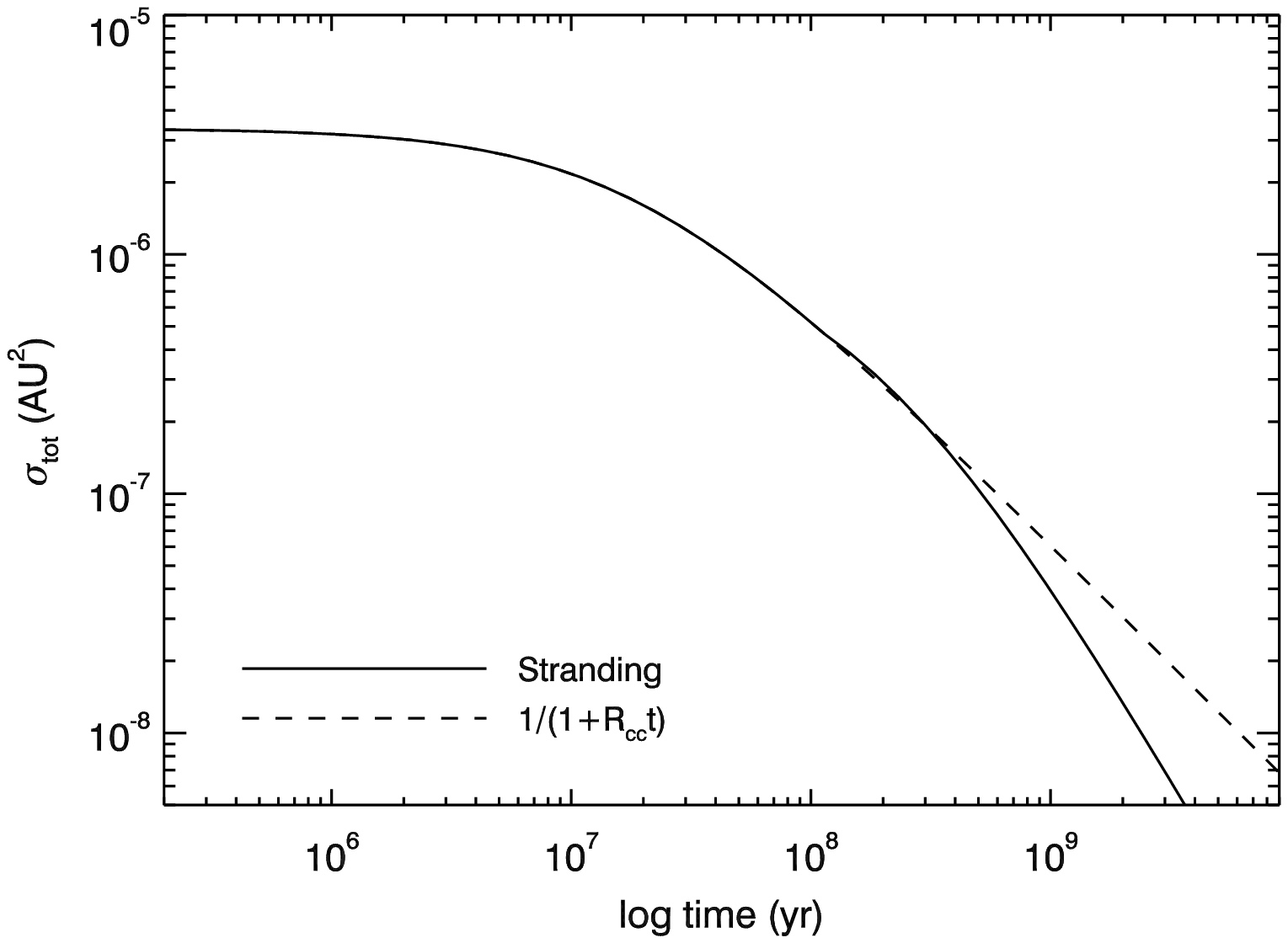,width=0.52\textwidth}
    \caption{Example evolution of the size distribution (left) and total surface area
      \stot\ (right) for 10\,Gyr. In the left panel the lines become lighter for later
      times and are logarithmically spaced in time. The dashed line shows stranded
      objects. In the right panel the evolution with stranding is shown as a solid line
      and the ``normal'' $1/t$ evolution with no stranding as a dashed
      line.}\label{fig:strandeg}
  \end{center}
\end{figure*}

If collisional evolution proceeds for long enough, the mass in the swarm will drop to the
point where it is comparable with the mass contained in a single largest object. Around
this time the evolution of the largest objects changes from being reasonably well
described by our particle-in-a-box formalism, to a regime where individual collisions and
cratering are important \citep{2010AJ....139..994B}. In this regime, the largest objects
are less likely to be destroyed due to their small number and the decreased number of
potential destructors. While our model cannot take cratering or stochastic collisions
into account, we can approximate the evolution by assuming that when the number of
largest objects drops too low, they lose their connection with the rest of the size
distribution and become ``stranded.'' The size of the largest non-stranded object is of
size \dc\ by definition, which decreases over time and leaves a relatively flat size
distribution of stranded objects between sizes \dc\ and \dmax, whose evolution is halted
due to a lack of would-be destructors.

We implement this simple approximation by assuming an object of size $D$ is stranded
when the number of objects between sizes $D/2$ and $D$
\begin{equation}\label{eq:nd2}
  n(D/2 \rightarrow D,t) = \frac{ K \left( 2^{3q_{\rm g}-3} -1 \right)
    ( 10^3 \, D )^{3-3q_{\rm g}} }{ 3 q_{\rm g} - 3 }
\end{equation}
drops to some number $N_{\rm str}$. This number sets the normalisation for the size
distribution of stranded objects. This assumption sets the size distribution slope of
stranded objects $q_{\rm str} = 1$ because $n(D/2 \rightarrow D,t)$ is independent of $D$
for $D_{\rm c} > D > D_{\rm max}$.

The first object is stranded at \tnl, which can be calculated from the initial number of
objects in this size range
\begin{equation}\label{eq:tnl}
  t_{\rm nleft} = \frac{ n(D_{\rm c}/2 \rightarrow D_{\rm c},t=0) }{ R_{\rm cc}(0) \, N_{\rm str} }
\end{equation}
The difference between $t_{\rm col}$ and \tnl\ is simply a measure of the initial number
$n$ of large objects, which take $n/N_{\rm str}$ collision times to become stranded. Like
the remaining mass once collisions occur, this time is independent of the initial cloud
mass.

At this point, \dc\ and \dmax\ become distinct sizes, with \dmax\ remaining fixed and
\dc\ decreasing with time as smaller objects are stranded. The remaining planetesimal
population decays at the collision rate for \dc\ size and strength objects and the mass
remaining in the size distribution below \dc. Therefore, by substituting \dc\ for \mtot\
in equation (\ref{eq:dmdt}), the evolution after \tnl\ obeys
\begin{equation}\label{eq:dcdt}
  \frac{dD_{\rm c}^3}{dt} = - D_{\rm c}^3 \, R_{\rm cc}
\end{equation}
where we have used $M_{\rm tot} \propto D_{\rm c}^3$ (for fixed \ndt). The collision rate
is for size \dc\ objects, on which both \mtot\ and \qd\ depend, and varies as $R_{\rm cc}
\propto D_{\rm c}^{1.2}$ (eqs. \ref{eq:qd} \& \ref{eq:rccdc}). Integrating equation
(\ref{eq:dcdt}) yields
\begin{equation}\label{eq:tnlevol}
  D_{\rm c} = \frac{ D_{\rm max} }{ \left( 1 + 0.4 (t - t_{\rm nleft}) / t_{\rm nleft}
    \right)^\alpha } \, .
\end{equation}
where $\alpha = 1/1.2$. This evolution is illustrated in Figure \ref{fig:strandeg}, which
shows the time evolution of the top section of the size distribution and \stot. The size
distribution initially decays straight down (i.e. $K$ decreases) with \dc\ fixed. When
there are only $N_{\rm str}$ largest objects left, \dc\ begins to decay as dictated by
Equation (\ref{eq:tnlevol}). The size distribution then moves to the left (smaller \dc),
with both \stot\ and \mtot\ continuing to decay. The right panel shows that \stot\ drops
more quickly after the first objects are stranded, tending to $\sigma_{\rm tot} \propto
t^{-1.75}$ (using eqs. \ref{eq:mtot} and \ref{eq:tnlevol}). Such evolution is potentially
interesting, as a dust cloud is accelerated towards being PR-dominated after stranding
due to the stronger decrease in small grains with time. Entering the PR-dominated regime,
the size distribution is effectively truncated at larger sizes, leading to an even faster
decay of \stot\ than shown in Figure \ref{fig:strandeg} \citep{2003ApJ...598..626D}.

This evolution is necessarily very simple because objects are only destroyed by
catastrophic collisions in our model. However, at these late stages the mass released
into the cascade by cratering may be as or more important
\citep{2010AJ....139..994B}. Therefore, though our model is physically plausible, the
actual evolution will depend on details such as the relative importance of catastrophic
disruptions vs. cratering or differences between pro and retrograde populations. We treat
$N_{\rm str}$ and $\alpha$ as free parameters when comparing our model with the Solar
System irregulars in \S \ref{ss:ss}.

\subsection{Observable quantities}\label{ss:obs}

Given the surface area in dust, we derive the flux density $F$ due to the cloud and
planet from both scattered light and thermal emission. Because distance and surface area
can be in different units (e.g. m, pc, AU), quantities in these equations have
dimensions. We take Solar System planetary effective temperatures and radii from
\citet{2000asqu.book.....C}. The stellar flux at the planet is
\begin{equation}
  F_\star = L_\star \, B_\nu(\lambda,T_\star) / ( 4 \, \sigma_{\rm K} \,
    T_\star^4 \, a_{\rm pl}^2 )
\end{equation}
where $\sigma_{\rm K}$ is the Stefan-Boltzmann constant. The scattered light seen from
Earth is \citep[e.g.][]{2002MNRAS.330..187C}
\begin{equation}
  F_{\rm scat} = F_\star \, A \, g \, Q / (\pi \, d_{\rm pl}^2)
\end{equation}
where $A$ is either the projected area of the planet or \stot\ for dust. The geometric
albedo $Q$ is assumed to be 0.08 for dust, similar to both the values for Kuiper belt
objects, Jovian trojans, and irregular satellites in the Solar System
\citep{2008ssbn.book..161S,2008DPS....40.6108M,2009AJ....138..240F} and to that inferred
for the Fomalhaut dust ring \citep{2005Natur.435.1067K}. For planets we use an albedo of
0.5. The phase function $g$ is set to unity for the Solar System because planets are on
exterior orbits. For extrasolar planets and swarms we set $g=0.32$, the value for a
Lambert sphere at maximum extension from the host star
\citep[e.g.][]{2002MNRAS.330..187C}. We use a blackbody estimate for the dust temperature
$T_{\rm dust} = 278.3 \, L_\star^{1/4} / a_{\rm pl}^{1/2}$\,K. The distance $d_{\rm pl}$
is the distance to the system in the extrasolar case, or $a_{\rm pl}$ in the Solar System
(i.e. roughly the distance for a Sun-Earth-Planet angle of 90$^\circ$ for outer
planets). For thermal emission
\begin{equation}
  F_{\rm th} = B_\nu(\lambda,T) \, A / d_{\rm pl}^2 \, .
\end{equation}
For dust we include a ``greybody'' decrease in emission of $210/\lambda$ beyond 210\,\um\
to account for inefficient emission from small grains
\citep[e.g.][]{2008ARA&A..46..339W}. Though the actual spectrum depends on the grain
properties and size distribution, this addition provides more realistic (sub-)mm flux
densities than a plain blackbody.

\section{Applications}\label{s:application}

In this section we apply our model to three different irregular satellite populations. We
first compare our model evolution with the Solar System's complement of irregulars and
make predictions of the current levels of dust. We suggest that the fate of dust changes
over time, with implications for regular satellites such as Callisto and Iapetus. We then
apply our model to possible satellite clouds around extrasolar planets. Finally, we apply
our model to Fomalhaut b.

\subsection{Solar System Irregular Satellites}\label{ss:ss}

\begin{figure*}
  \begin{center}
    \vspace{-0.2in}
    \hspace{-0.35in} \psfig{figure=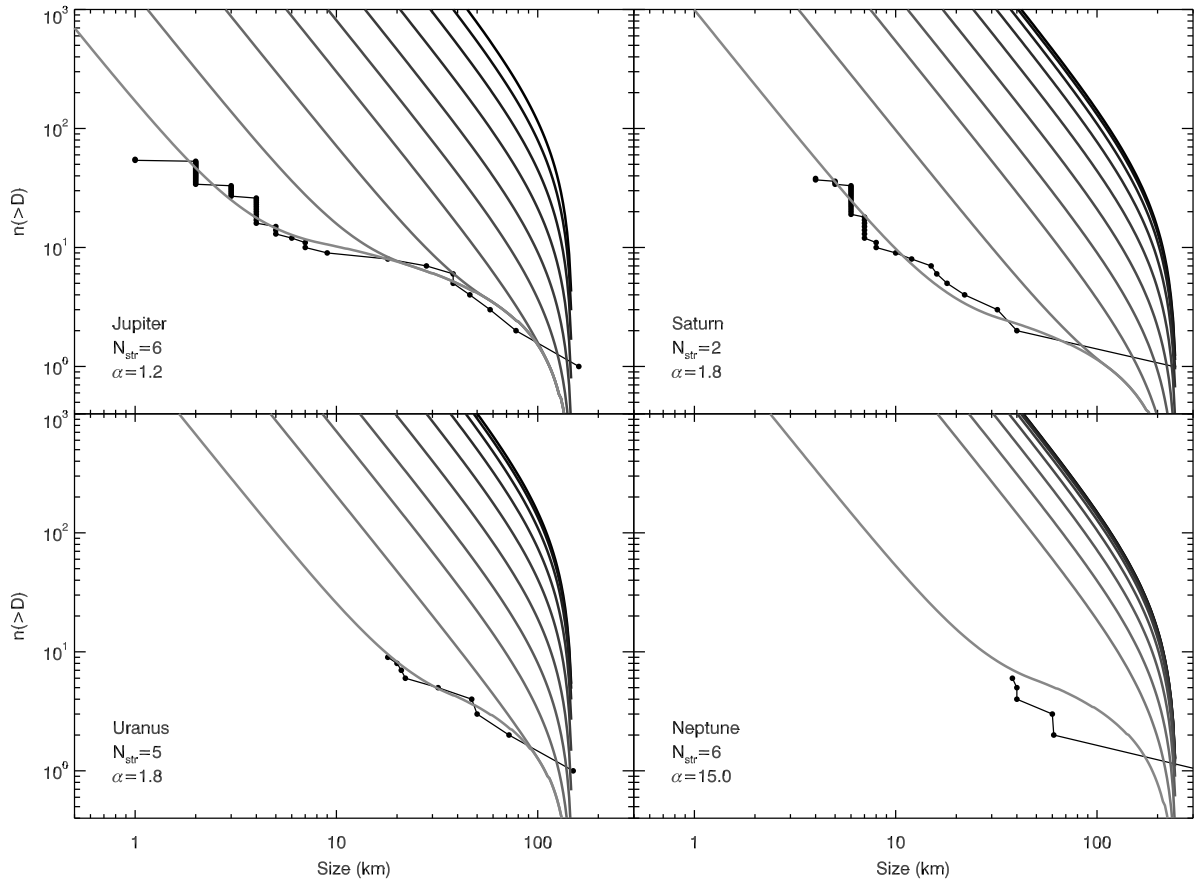,width=1\textwidth}
    \caption{Comparison of our model (thick curves) with the Solar System's irregular
      satellites (dots and thin lines, pro/retrograde populations have been added
      together). Curves show ten logarithmically spaced times between $10^6$ and $4.5
      \times 10^9$\,years (dark to light). The legend indicates the planet and values for
      $N_{\rm str}$ and $\alpha$. Model curves turn down at the largest sizes because
      they are derived from integrating $n(D)$ (and there are zero objects larger than
      \dmax).}\label{fig:ssirreg}
  \end{center}
\end{figure*}

In this section we compare our model with the Solar System's irregular satellites. We
model a swarm with initial mass $M_{\rm tot}(0) = 0.01\,M_{\rm moon}$. This mass ensures
all swarms are collision limited and therefore does not influence the model. We set
$D_{\rm c} = 150$\,km for Jupiter and Uranus and $D_{\rm c} = 250$\,km for Saturn and
Neptune. We set the density similar to known irregular satellites $\rho = 1500$\,kg
m$^{-3}$ \citep[e.g.][]{2007ARA&A..45..261J}. Though we assume $f_{v_{\rm rel}} = 4/\pi$
at each planet, this number varies somewhat depending on the specific orbital properties
of each swarm. We use the average $\eta$ of known irregulars at each planet.

Figure \ref{fig:ssirreg} shows a comparison of our model with the Solar System's
irregular satellites.\footnote{Taken from
  \href{http://www.dtm.ciw.edu/users/sheppard/satellites/}{http://www.dtm.ciw.edu/users/sheppard/satellites/}
  in March 2010} We have combined the pro/retrograde satellites for this
distribution. The simplicity of our model means that we cannot account for the different
evolution of pro/retrograde populations, which appears to be important at Uranus
\citep{2010AJ....139..994B}. Because all satellites at each planet are at a similar
distance from Earth, these distributions are near-complete to the smallest observed size
and need no correction \citep[e.g.][]{2005AJ....129..518S}.

We vary the normalisation $N_{\rm str}$ and rate of decay after stranding power-law index
$\alpha$ to obtain a by-eye fit. Variation of these parameters over a fairly small range
allows an excellent match for Jupiter, Saturn, and Uranus. Given the differences in the
irregular populations at each planet and the simplicity of the prescription for
stranding, some variation is expected. The first few columns of Table \ref{t:ssdust} show
the $\eta$, $t_{\rm col}$, and \tnl\ for each planet. The slower collisional evolution
for more distant planets means that \tnl\ is longer and that the size distribution at the
current epoch turns up (i.e is stranded) at larger sizes. The minimum known satellite
size increases for more distant planets, making the model comparison less certain for
Uranus and Neptune. We do not know if their size distributions are similar to Jupiter and
relatively flat to $\sim$8\,km, or instead turn up at larger sizes like our model. If
evolution at Uranus is similar to Jupiter and Saturn, our model predicts that the size
distribution should turn up at a few tens of km due to the slower collisional evolution.

There are few Neptune irregulars to compare with, perhaps because many were depleted by
Nereid \citep{2003AJ....126..398N} and/or Triton
\citep{1989Sci...245..500G,2005ApJ...626L.113C}.  Nereid's low orbital inclination
(7$^\circ$ relative to the ecliptic) and relatively close prograde orbit ($\eta = 0.05$),
as well as a colour and albedo similar to the Uranian satellites Umbriel and Oberon
\citep{1997Icar..126..225B}, mean that it may in fact have formed as a regular satellite
of Neptune. For these reasons \citet{2010AJ....139..994B} did not model Neptunian
irregulars.

Given these complications, it is perhaps unsurprising that our model for Neptune's
irregulars needs very different \nstr\ and $\alpha$ to the other giant planets, and is
still a poor match. This difference is due to the slow collisional evolution, which
predicts that \tnl\ is a sizeable fraction of the Solar System age. With stranding
occurring at such a late time, the subsequent evolution must be very rapid to deplete the
population to that currently observed. With values for \nstr\ and $\alpha$ more like the
other three planets, our model predicts that Neptune would have two orders of magnitude
more satellites (i.e. be at about the second to lowest curve). This discrepancy suggests
that either the initial conditions for Neptune's swarm were quite different to the other
planets, or as already proposed, the irregulars were affected by Nereid and Triton.

Based on these comparisons we conclude that our model provides a reasonable description
of the collisional evolution of an irregular satellite swarm.

\subsubsection{Current irregular dust levels}\label{ss:ssdust}

Given the ongoing collisional erosion of the irregular satellites, the presence of dust
is inevitable. Here we make some estimates of the expected level of dust at each planet
and relate them to a few relevant observations. Table \ref{t:ssdust} shows estimates for
each planet based on \stot\ from the models of Figure \ref{fig:ssirreg} at $t = 4.5
\times 10^9$\,years. That is, they are extrapolated using our size distribution and
independent of the collisional model. We calculate the flux density and surface
brightness at 1 and 100\,\um\ (distributed uniformly over a disk of radius 0.5\rh\ for
simplicity), which roughly correspond to peaks in scattered light and thermal emission
respectively.

\begin{table*}
  \begin{center}
    \begin{tabular}{lrrrrrrrlrrlrrr}
      \hline
      \vspace{2pt}
      & & & & & & \stot & & & \multicolumn{3}{c}{$\underline{\phantom{awwwww} 1
\,\mu {\rm m} \phantom{awwwww}}$} &
      \multicolumn{3}{c}{$\underline{\phantom{aawwww} 100\,\mu {\rm m}
          \phantom{aawwww}}$} \\
      & \rh & $\eta$ & $t_{\rm col}$ & \tnl & \dmin & ($10^{-9}$ & dens & \chipr & $F_{\rm dust}$ &
      $F_{\rm pl}$ & $B$ & $F_{\rm dust}$ & $F_{\rm pl}$ & $B$ \\
      Planet & (AU) & & (Myr) & (Myr) & (\um) & AU$^2$) & (km$^{-3}$) & & (Jy) & (Jy) & (MJy/sr) & (Jy) & (Jy) & (MJy/sr) \\
     \hline
Jupiter & 0.35 & 0.4 & 2.5 & 170 & 12 & 1.5 & 0.9 & \phantom{1}0.05 & 12.0 & 37000 & 
0.0034 & 960 & 450000 &0.26 \\
Saturn & 0.43 & 0.3 & 17.0 & 740 & 16 & 7.5 & 1.5 & \phantom{1}0.8 & 5.7 & 2400 & 0.0035
 & 840 & 60000 &0.52 \\
Uranus & 0.46 & 0.2 & 14.0 & 1100 & 24 & 16.0 & 1.1 & 13.0 & 0.8 & 28 & 0.0016 & 
210 & 950 &0.45 \\
Neptune & 0.77 & 0.2 & 250.0 & 3500 & 23 & 36.0 & 0.6 & 13.0 & 0.3 & 4 & 0.0005
 & 99 & 350 &0.19 \\
      \hline
  \end{tabular}
  \end{center}
  \caption{Irregular satellite model and dust properties at each planet. Estimates
    are based on the models in Figure \ref{fig:ssirreg}. The last six columns show
    estimates of the total dust cloud and planet flux density at opposition, and the surface
    brightness if the cloud were evenly spread over a disk with radius 0.5\rh, at 1 and
    100\,\um. The peak surface brightness is $\sim$6 times higher than shown here
    \emph{if} grains are distributed as in Figure \ref{fig:saturnpic}.}\label{t:ssdust}
\end{table*}

The predicted surface brightness levels are much fainter than the background in the
ecliptic, but expected to vary on a similar scale (i.e. degrees). For comparison, the
Zodiacal background at 1.25 and 100\,\um\ is about 0.4 and 9\,MJy/sr respectively
\citep{1998ApJ...508...44K}. Detection of these clouds therefore requires accurate
subtraction of this (and the galactic and cosmic) background.

Of course dust that originates from irregular satellites has been detected already (see
also \S \ref{ss:ssdisc} below). The largest of Saturn's rings is probably fed by material
generated when inter/circumplanetary grains impact the irregular satellite Phoebe
\citep{2009Natur.461.1098V}. At 24\,\um, this ring has a surface brightness of
$\sim$$0.4$\,MJy sr$^{-1}$, which is less than 1\% of the ($\sim$70\,MJy/sr) zodiacal
background. For comparison, our Saturn swarm model has $\sim$0.2\,MJy sr$^{-1}$ when
spread over a 0.5\,\rh\ radius disk.

\begin{figure}
  \begin{center}
    \hspace{-0.35in} \psfig{figure=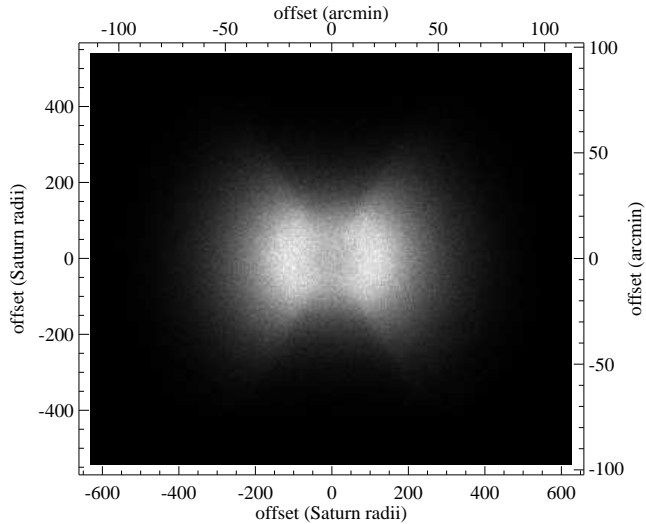,width=0.52\textwidth}
    \caption{Simple model of Saturn's irregular satellite dust cloud at 24\,\um. Grains
      follow parent body orbits. The total flux is 320 Jy and the scale is a linear
      stretch between 0 and 1.25 MJy/sr. Flux from Saturn itself is not
      included.}\label{fig:saturnpic}
  \end{center}
\end{figure}

To take a more detailed look at Saturn, Figure \ref{fig:saturnpic} shows an example of
what such a cloud might look like from Earth at 24\,\um. In creating this image we have
assumed that the dust follows the same orbits as the parent bodies, which have semi-major
axes distributed between 0.04--0.16\,AU, and $e = 0.1$--0.6 based on Figure 7 of
\citet{2007AJ....133.1962N}. The image would therefore look the same at any wavelength,
though the total flux changes. The total flux in this 24\,\um\ image is 320 Jy, which
yields a peak surface brightness of around 1.25 MJy/sr. Therefore, the surface brightness
predictions in Table \ref{t:ssdust} are a factor $\sim$6 too low if grains follow their
parent body orbits. The peak level is about three times brighter than Saturn's Phoebe
ring.

Though the cloud is spread over a much larger region of sky, for it to have evaded
detection thus far, particularly in the Phoebe ring observations, it appears that our
prediction is at least a few times too high. However, \citet{2009Natur.461.1098V} note an
apparent trend of increasing surface brightness towards the Phoebe ring, which could be
due to dust from irregular satellites. The narrow vertical range of their Figure 3 does
not constrain our prediction, because the Phoebe ring could be sitting on top of a larger
background.

Several uncertainties with our model may explain this apparently high prediction for the
level of dust at Saturn, the most likely being that small departures from our assumed
size distribution can lead to large differences in the predicted surface area in dust. In
addition, we have assumed that grains follow the orbits of their parent bodies, but
\citet{2002Icar..157..436K} show that the detailed cloud structure is more
complex. Finally, Table \ref{t:ssdust} shows that grains around Saturn may be in the
PR-dominated regime, which can alter the cloud structure and reduce the dust level.

Despite our apparent over-prediction of the cloud surface brightness, detection of such a
cloud is difficult, not least because it may be unexpected. The large extent means
achieving sufficient coverage (ideally the entire Hill sphere) is expensive at IR
wavelengths (i.e. using space telescopes). Poorly characterised dust bands will also
hinder background subtraction. At longer wavelengths, such as those covered by
\emph{Herschel} SPIRE, achieving decent coverage is easier, but the background comprises
a combination of Zodiacal and galactic light. Care would be needed to ensure good
background subtraction that minimises elongation effects.

Now looking at Jupiter, from Galileo dust detection data \citet{2002Icar..157..436K}
derive a constant dust number density of $\sim$10\,km$^{-3}$ between 50 and
350\,\rj. \citet{2010P&SS...58..965K} report detection of micron-size particles (and note
the lack of detections of smaller particles) by the Galileo dust detector at a distance
of approximately 350\,\rj. Such large planetocentric distances are prime irregular
satellite territory, being about half the Hill radius. These grains have previously been
explained as ejecta from impacts of interplanetary dust on irregular satellites, which
produce about the same level of dust as detected by Galileo
\citep{2002Icar..157..436K}. The predicted space density of dust for Jupiter shown in
Table \ref{t:ssdust}\ (again within 0.5\,\rh) is at a lower level to the Galileo
detections and may not have contributed to the measurements. However, the space density
is an extremely strong function of \dmin. For example, if the minimum grain size in our
model were 1\,\um\ the predicted space density at Jupiter would be
$\sim$100\,km$^{-3}$. Therefore, Galileo observations constrain either the grain size in
our model to be larger than a few microns, or the dust level to be lower than predicted
if the minimum grain size is 1\,\um.

\subsubsection{Fate of Dust}\label{ss:ssdisc}

The best way to probe the small end of the irregular satellite size distribution is to
detect the dust cloud directly. However, as noted above such an observation is difficult
and there are complementary ways to detect irregular dust. One of the most interesting
signatures exists on the surfaces of some regular satellites. The orbits of the smallest
grains are strongly affected by radiation forces and may end up on a regular
satellite. However, because this deposition is only one of several possible fates a grain
may meet, an understanding of which is more likely (and when) is needed to make a strong
connection between irregular dust and regular satellite surfaces. Our model is too simple
to model irregular satellite evolution in the Solar System at a detailed level. However,
we offer some order of magnitude arguments concerning the fate of dust, which highlight
questions that should be asked by more detailed studies.

Like a circumstellar disk, a young circumplanetary swarm will most likely be collision
dominated. At this stage grains are lost to the planet and interplanetary space. When the
mass has been sufficiently depleted it becomes PR dominated and grains spiral in toward
the planet. Grains destined to impact the planet, or pass nearby, may meet a third fate
and be swept up by a regular satellites. Though only a small fraction of mass may be lost
this way, it is important due to the visible effect of leading/trailing asymmetries on
some tidally locked Solar System regular satellites, such as Callisto and Iapetus. These
asymmetries are thought to arise from the higher accretion rate of retrograde dust by the
leading hemisphere \citep[e.g.][]{1996ASPC..104..179B}.

Though most grains in collision dominated swarms either leave the Hill sphere or impact
the planet due to radiation pressure, nearly all must pass through the regular satellite
domain to do so. Grains destined to hit the planet or leave the Hill sphere do not reach
their maximum eccentricity immediately, but instead make a number of pericenter passages
first. The smallest grains with $\eta \sim 0.1$--0.5 only complete a few, to a few tens
of orbits before their eccentricities exceed unity (i.e. $t_{\rm per,grain} \sim
\eta^{3/2} \, t_{\rm per,planet}$) and are therefore unlikely to encounter regular
satellites before removal.

Grains in the small size range where $e$ grows large enough to pass through the regular
satellite region but not high enough to hit the planet or leave the Hill sphere have the
best chance of colliding with regular satellites. However, this collision time must be
shorter than the time for grains to collide with themselves. Taking Saturn as an example,
at $\eta=0.3$ with $\sigma_{\rm tot} = 7.5 \times 10^{-9}$\,AU$^2$ (Table \ref{t:ssdust})
the current time for collisions between the smallest grains is predicted to be
$\sim$10$^6$\,years (eq. \ref{eq:tcoldust}). In the past the level of dust was much
higher, and the grain-grain collision time correspondingly shorter.

Continuing with the Saturn example, grains with $0.83 > e > 0.997$ traverse the region
between Iapetus' orbit and Saturn's surface near pericenter. This range corresponds to
grains of sizes $16 > D > 25$\,\um. The orbital period $t_{\rm per}$ is about 3\,years
for these grains. Grains on coplanar orbits with Iapetus in this size range spend no more
than 40\% of their lifetimes in Iapetus-crossing orbits. Iapetus occupies about
1/20000\,th of its orbital torus, so coplanar grains have a collision time of
$\sim$$10^4$ orbits, or $\sim$$10^4$ years. This time is shorter than the grain-grain
collision time, so coplanar grains in this size range appear likely to impact Iapetus.

However, grains on inclined orbits (i.e. the majority) only have a chance of hitting
Iapetus if they cross Iapetus' plane at precisely the right radial distance. Iapetus is
$\sim$1500\,km across, compared to a semi-major axis of $3.5 \times 10^6$\,km, so the
chances of a plane-crossing particle encountering Iapetus \emph{orbit} are roughly
1/2000. The chance of an inclined grain impacting Iapetus at each plane crossing is
$\sim$10$^{-7}$ per orbit, or an impact time of $\sim$10$^{7}$ years. The Iapetus
collision time is therefore longer than the grain-grain collision time, so grains are
more likely fragmented to smaller sizes first. Taking these numbers as representative, it
appears that in collision dominated swarms only a small fraction of grains, those within
a small size range and on coplanar orbits, will impact regular satellites.

In contrast to the highly variable eccentricities of radiation pressure affected grains,
PR drag causes grain orbits to collapse slowly. The chance of impacting a regular
satellite is larger than for the more distant and eccentric radiation pressure induced
orbits because grains orbit the planet more often as their semi-major axes shrink. For
example, in the case of grains released from Phoebe, collision with Iapetus is nearly
guaranteed, with escapees destined to hit Hyperion or Titan
\citep[e.g.][]{1996ASPC..104..179B}.

Table \ref{t:ssdust} shows that dust from irregular satellites at Jupiter and Saturn is
near or in the regime where PR drag dominates, but dust at Uranus and Neptune is not. If
grains in PR dominated swarms are more likely to impact regular satellites, this finding
is consistent with the brightness asymmetries seen on Callisto and Iapetus, which are
less marked (but still present) on the Uranian satellites.

While we suggest that the Jupiter and Saturn swarms are PR dominated now, they were
collision dominated in the past. The possibility that collision dominated swarms do not
coat regular satellites as efficiently as PR dominated ones therefore has implications
for the interpretation of the brightness asymmetries. If the mass deposited on regular
satellites is some fraction of the total mass lost, this fraction will increase as the
swarm changes from collision to PR dominated. This change will affect the accretion
history and more mass may be accreted at late stages if the difference in accretion
efficiency is large enough.

Such interpretations may be too simple. For example, Iapetus' asymmetry is likely due to
accretion of grains from Saturn's Phoebe ring, so is due to grains released from an
individual irregular satellite rather than the cloud in general. In summary, further
study of the fate of irregular satellite debris and potential observables requires
consideration of the competition between collisions, PR drag, and radiation pressure.

\subsection{Irregular Satellite Clouds around Exoplanets}\label{ss:exo}

While the dust produced by Solar System irregular satellites is at very low levels, these
clouds were many times brighter at earlier times. In this section we explore the
prospects for discovery of young extrasolar circumplanetary swarms. Like circumstellar
debris disks, dust can be discovered in scattered light at optical wavelengths, and
thermal emission at IR wavelengths. In the case of thermal emission, one may need to
distinguish a planetary atmosphere from dust at a similar temperature via spectral
features.

Because there is little \emph{a priori} reason to choose any particular system
configuration, we first model a particular system and then show which configurations are
detectable in a more general sense. Clearly, to be detectable with current or near-future
technology these systems must possess more dust than predicted for the Solar System's
irregulars. Further, because the first objects discovered where sensitivity is an issue
tend to be the most extreme (e.g. hot Jupiters), we do not restrict the initial swarm
masses. While the initial conditions used by \citet{2010AJ....139..994B} based on the
\citet{2007AJ....133.1962N} simulations were at most a small fraction of the Moon's mass,
extrasolar systems may have much more massive planetesimal belts from which to capture
satellites, and mechanisms disfavoured for the Solar System may also operate. We discuss
capture mechanisms further in \S \ref{ss:exodisc}.

In this section we first use the example of a Jupiter-mass planet orbiting an A5 star at
10\,pc. We choose this spectral type of star simply because the only planets directly
imaged to date orbit A-stars
\citep{2008Sci...322.1345K,2008Sci...322.1348M,2009A&A...493L..21L}. However, more
massive stars do not necessarily produce brighter clouds because their high luminosity
and mass increases \dmin\ and changes the collision rate. For the evolution of the
planet, we use the non-irradiated Jupiter-mass ($Z=0.02$) model from
\citet{2008A&A...482..315B}. We set $M_{\rm tot}(0) = 10\,M_{\rm moon}$, and $D_{\rm c} =
250$\,km. Another choice is when the satellite swarm is captured (or otherwise formed)
and evolution begins. For simplicity, we assume swarms start evolving at $t=0$.

\begin{figure}
  \begin{center}
    \vspace{-0.2in}
    \hspace{-0.35in} \psfig{figure=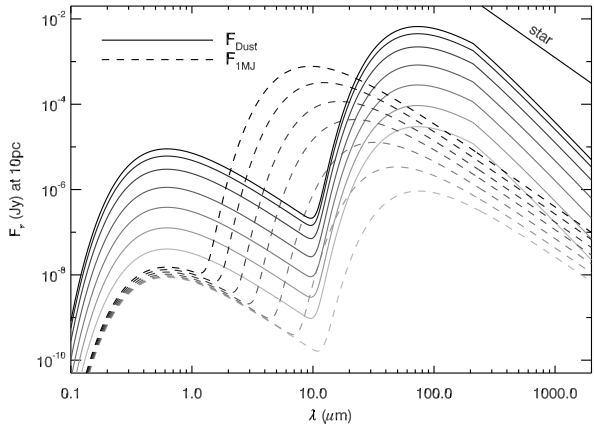,width=0.52\textwidth}
    \caption{Example satellite swarm evolution around a 1\,\mj\ extrasolar planet at
      50\,AU around an A5 star between $10^7$ (darkest curves) and $10^{10}$\,years
      (lightest curves, logarithmically spaced). A small part of the stellar spectrum is
      visible.}\label{fig:egpspec}
  \end{center}
\end{figure}

Figure \ref{fig:egpspec} shows the evolution of scattered light and thermal emission from
both planet and dust cloud when the planet orbits at 50\,AU. The Figure is drawn to
highlight where the spectrum of each component moves over time (which is downwards in
this plot). In scattered light, the dust cloud is initially much brighter than the
planet, but decreases significantly as the satellite swarm collides and the dust level
subsides. The planet shrinks somewhat as it cools, but in scattered light is nearly
constant. At early times the planet's thermal emission peaks slightly short of 10\,\um,
but moves to longer wavelengths over time. Thermal emission from the dust beyond 20\,\um\
is brighter than the planet at all times. They become comparable at mm wavelengths at
late times, which shows the importance of including grain emission inefficiencies.

\begin{figure}
  \begin{center}
    \vspace{-0.2in}
    \hspace{-0.35in} \psfig{figure=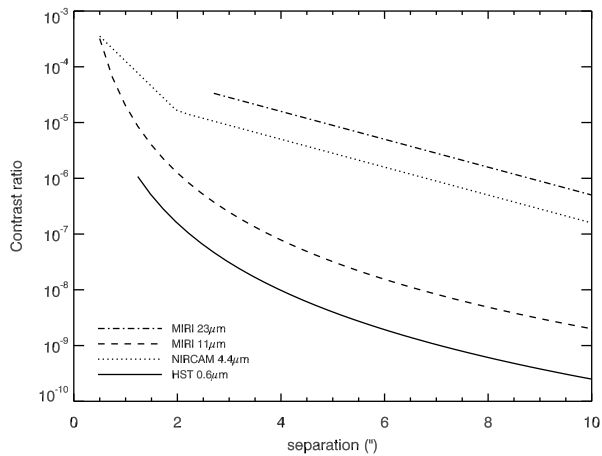,width=0.52\textwidth}\\
    \caption{Adopted instrument contrast ratios.}\label{fig:egpcontr}
  \end{center}
\end{figure}

This evolution therefore highlights wavelengths where irregular satellite clouds may be
bright enough to be detected, both in absolute terms and relative to thermal and
scattered emission from the planet. Unsurprisingly these lie at the thermal and stellar
peaks. The key to detection lies with rejection of starlight, the same issue faced by
those looking for planets or circumstellar debris disks
\citep[e.g.][]{2007prpl.conf..915B,2010PASP..122..162B}. This rejection is characterised
by the star/planet contrast ratio that a particular instrument can detect, which is
usually a function of angular separation between the two. While Figure \ref{fig:egpspec}
shows that the youngest dust clouds may be bright enough for detection by the
\emph{Herschel} Photodetector Array Camera and Spectrometer (PACS) at 100\,\um, the 7''
resolution means that resolving these systems is difficult. We therefore focus on optical
and near/mid-IR wavelengths. We model the detectability of satellite swarms using simple
approximations to published contrast ratios for several instruments.

For the \emph{Hubble Space Telescope Advanced Camera for Surveys} (HST ACS) the contrast
is based on actual roll-subtracted coronagraphic observations \citep{Krist2006} and we
set an absolute detection limit of 0.1\,$\mu$Jy. We use predicted \emph{James Webb Space
  Telescope} (JWST) Near IR Camera (NIRCAM) 4.4\,\um\ and Mid IR Imager (MIRI) 11.4
contrast ratios from \citet{2010PASP..122..162B}. For MIRI at 23\,\um, we use the same
contrast as at 11.4\,\um\ for the same $\lambda/D$ \citep[i.e. at twice the
separation,][]{2004EAS....12..195B}. The other difference from the three shorter MIRI
coronagraph wavelengths is that only a Lyot stop is offered at 23\,\um\ so the inner
working angle is larger. We set absolute detection limits of 68\,nJy, 2.5\,$\mu$Jy, and
50\,$\mu$Jy for 4.4, 11.4, and 23\,\um\ respectively. These sensitivities are based on a
5\,$\sigma$ detection in the difference of two 1\,hour exposures
\citep{2010PASP..122..162B}. The contrast ratios are shown in Figure \ref{fig:egpcontr}.

\begin{figure}
  \begin{center}
    \vspace{-0.2in}
   \hspace{-0.35in} \psfig{figure=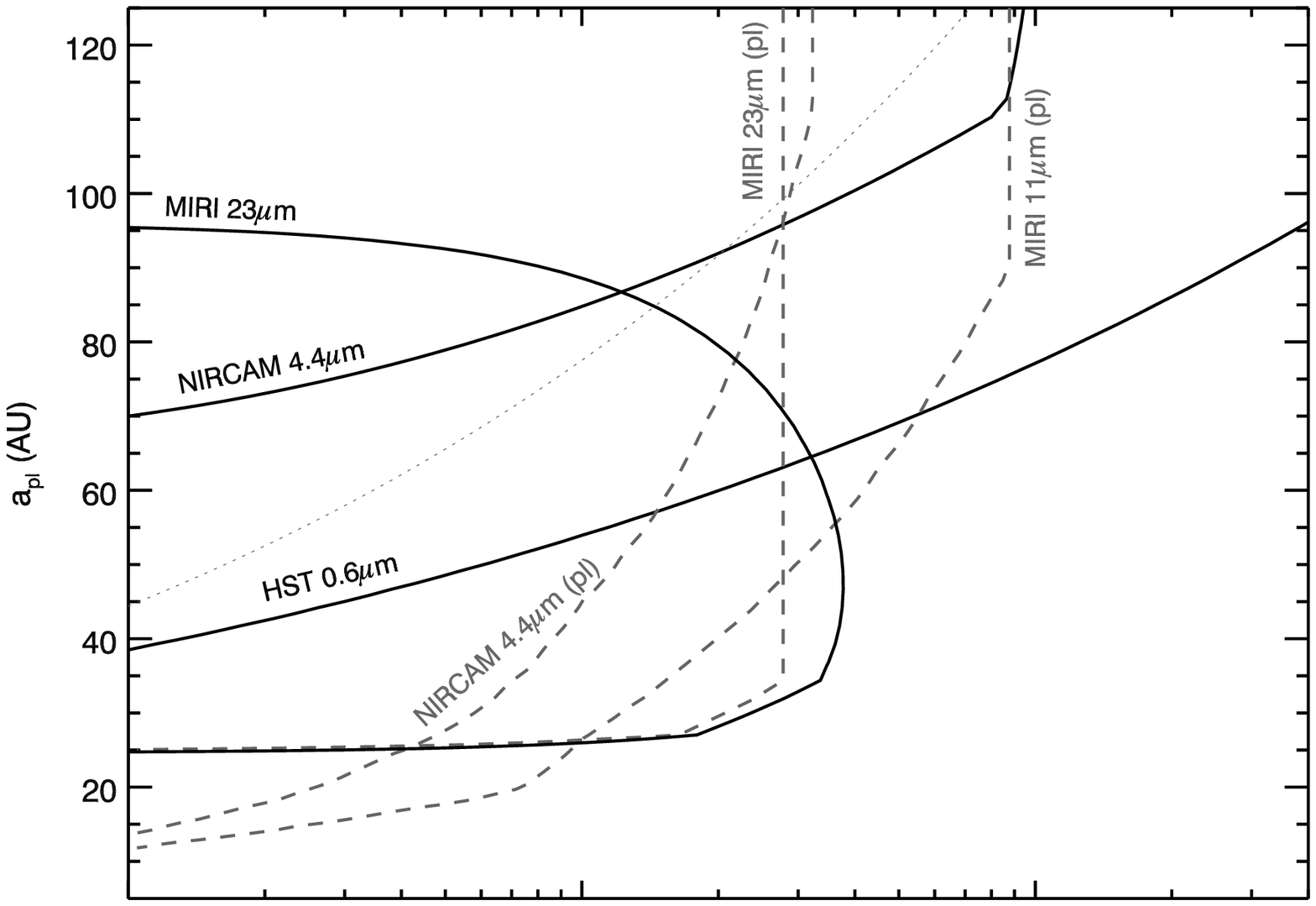,width=0.52\textwidth}\\
   \vspace{-1.2cm}
   \hspace{-0.35in} \psfig{figure=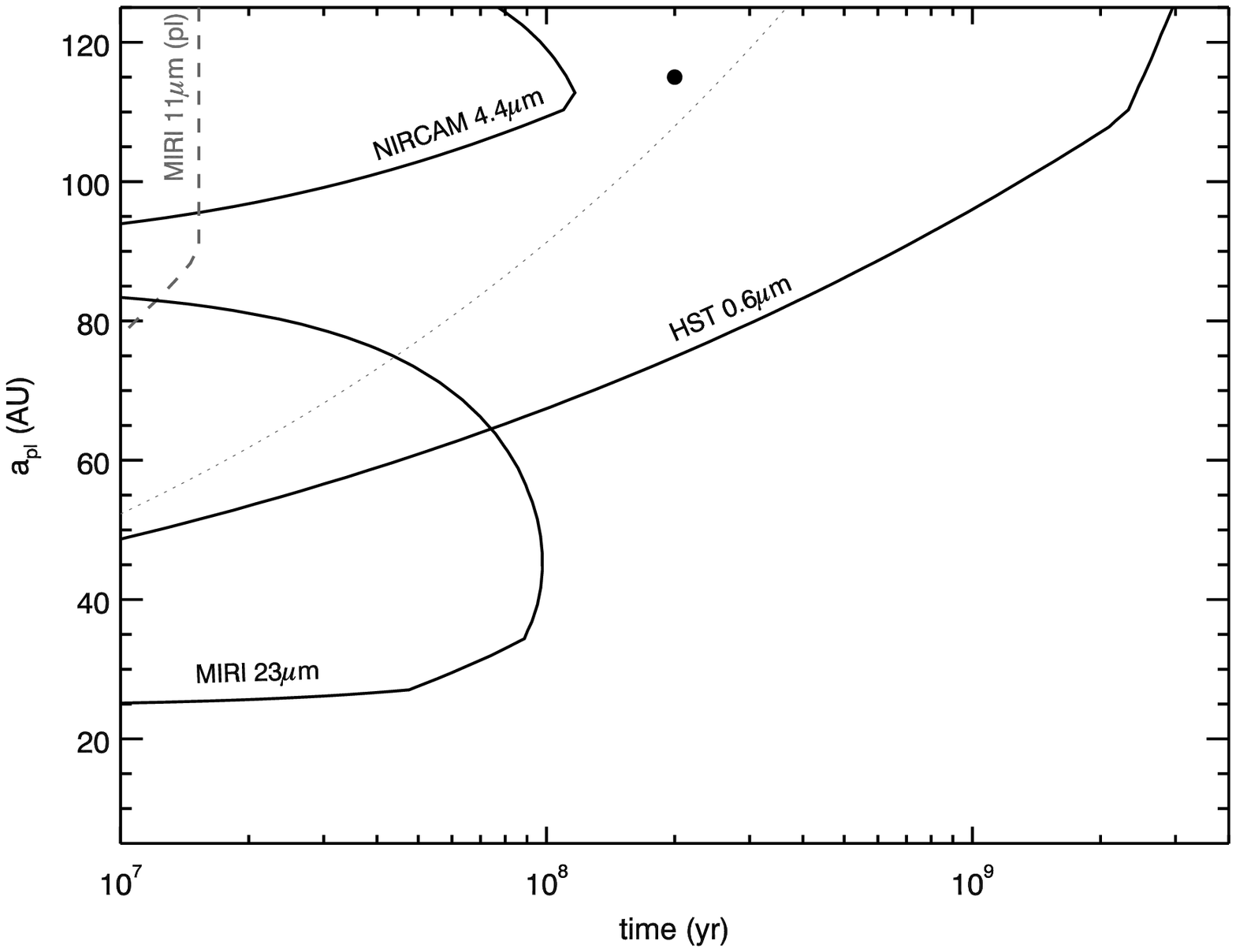,width=0.52\textwidth}\\
   \caption{Regions in \apl\ vs. time space where satellite swarms (dark lines) and their
     host planets (grey dashed lines) can be detected with various instruments. All space
     to the left of each contour is detectable by that instrument (swarms and planets
     become fainter as they move right). The dashed line shows \aopt
     (eq. \ref{eq:aopt}). The top panel shows detection contours for a Jupiter-mass host,
     and the lower panel contours for a $20\,M_\oplus$ planet. Fomalhaut b is marked by a
     filled circle (see \S \ref{ss:fomb}).}\label{fig:egpcoron}
  \end{center}
\end{figure}

Figure \ref{fig:egpcoron} shows detectability contours for the Jupiter-mass planet at a
range of possible planetary semi-major axes from our A5 star at 10\,pc. For each
instrument there are two curves; one for the swarm (black curve) and one for the planet
(grey curve). Swarms and planets in the space to the left of curves are detectable. The
optimum planet semi-major axis for detection in scattered light for our chosen parameters
(eq. \ref{eq:aopt}) is drawn as a dotted line.

Looking at the detection limits for each instrument individually, HST ACS detects
scattered light from swarms across a wide range of the parameter space. As expected from
equation (\ref{eq:aopt}), these swarms lie many tens of AU from the star. Detectable
swarms are at larger semi-major axes at later times as swarms around planets on closer
orbits deplete. As expected from Figure \ref{fig:egpspec}, Jupiter-mass planets are hard
to detect with ACS at any separation (so there is no grey HST curve on
Fig. \ref{fig:egpcoron}).

JWST instruments are well suited to planet detection by design so planets and swarms are
detectable. The detection space for satellite swarms with NIRCAM covers semi-major axes
greater than 70-120\,AU until a Gyr. Planets are detectable for a shorter time but to
much closer separations. The difference is because the swarms are detected in scattered
light, and planets from their thermal emission. At 11\,\um\ MIRI does not detect the
satellite swarms, but is ideal for detecting cooling planets, whose SEDs initially peak
in or near this region. The detection region therefore covers even more age and
separation space than at 4.4\,\um.

At 23\,\um\ MIRI detects the cooling planet, but also sees the Wein side of the dust
cloud for systems up to $3 \times 10^8$\,years old. At this wavelength the visibility
space for MIRI lies closer than suggested by \aopt\ because the dust clouds are
hotter. With the exception of the vertical parts of curves (most notably at 23\,\um),
detectability in this example is set by contrast rather than absolute sensitivity.

Therefore, optical wavelengths appear to be the best place to detect extrasolar irregular
satellite swarms around Jupiter-mass planets. Though detectable, at IR wavelengths the
swarms are fainter than their host planets. Discovery then involves finding whether a
(presumably previously known) planet has an optical or IR excess.

Over the parameter space in Figure \ref{fig:egpcoron} we find that the largest excess
ratio ($F_{\rm swarm}/F_\star$) is about 2\% at 100\,\um\ and 3\% at 160\,\um\ for a
young swarm at 70--80\,AU. These excesses would not be detectable given current
uncertainties with \emph{Herschel} PACS photometry \citep{2010A&A...518L...2P} and planet
atmospheres. Given the resolution, a sufficiently bright swarm detected this way would be
classed as a circumstellar disk unless it were resolved at shorter wavelengths or
orbiting a nearby star. Such a limitation may be overcome by high resolution facilities
such as ALMA.

The predictions of Figure \ref{fig:egpcoron} for all instruments can be extended in
several directions. For the same star moved to 5\,pc, swarms and planets at the same
semi-major axis are more detectable because they lie at larger angular separations (where
the instrument contrast ratio is better). Alternatively, a swarm around a planet at 10''
separation around a star at 10\,pc (planet at 100\,AU) is about equally detectable at
optical/near-IR wavelengths when the star is at 5\,pc (planet at 50\,AU), but more easily
detectable at mid-IR wavelengths because the dust is hotter. For the same range of planet
semi-major axes shown in Figure \ref{fig:egpcoron}, the same swarms become much harder to
detect beyond a few tens of parsecs. The difficulty arises due to the poorer contrast at
smaller angular separations. If swarms are detectable beyond a few tens of parsecs, they
must have of order an Earth-mass or more in irregular satellites.

As the stellar mass decreases the detection space at optical/near-IR wavelengths
increases due to the much lower stellar luminosity and consequently smaller minimum grain
size. However, the clouds become undetectable in the mid-IR because the Wien side of
their spectrum is too faint. The planets themselves become somewhat more detectable at IR
wavelengths because their thermal emission is the same but the star is fainter (and we
have assumed all planetary luminosity is intrinsic).

Decreasing the planet mass can make for undetectable planets that host detectable
swarms. The weak Hill radius dependence on planet mass means that collisional evolution
is only somewhat slower (eq. \ref{eq:rccdc}). In addition, the minimum grain size only
depends weakly on planet mass (eq. \ref{eq:dmin}). To highlight the relative unimportance
of planet mass on the evolution of satellite swarms, the lower panel of Figure
\ref{fig:egpcoron} shows detectability contours for a $20\,M_\oplus$ planet. The much
fainter planet can only be detected at the earliest times with MIRI at 11\,\um. Despite
the 16 times decease in planet mass, the HST detection space is not much smaller than for
the Jupiter-mass planet in the upper panel and still covers a wide range of orbits and
ages. The detection space for NIRCAM at 4.4\,\um\ is extremely small. At 23\,\um, MIRI
can still detect young swarms orbiting planets at 25 to 80\,AU from the star. This Figure
therefore shows that relatively low-mass planets that would otherwise be invisible can be
detected thanks to the luminosity of their irregular satellite swarms. As we argue below
for the specific case of Fomalhaut b, these swarms could be misidentified as more massive
planets in the first instance.

\subsubsection{Discussion}\label{ss:exodisc}

The ability of planets to capture irregular satellite swarms and reside at tens of AU
sets the likelihood that any will be detected outside the Solar System. The wide range of
planet-masses about which swarms should be detectable means that how capture mechanisms
and efficiencies change with planet mass is important. In particular, our prediction that
swarms may be visible around relatively low-mass planets relies on the ability of these
planets to capture swarms.

Giants forming by core accretion may capture irregulars passing through their primordial
envelopes \citep[e.g.][]{1979Icar...37..587P}, or by ``pull-down'' during a phase of
rapid growth \citep[e.g.][]{1977Icar...30..385H}. Giants that form by gravitational
instability may have analogous capture mechanisms. \citet{2007ARA&A..45..261J} suggest
that the similarity of the irregular populations at each planet argues against gas giant
specific capture processes for the Solar System. However, \citet{2010AJ....139..994B}
note that because the size distributions are a result of the collisional evolution, they
cannot be used to constrain the capture mechanism.

Any planet may capture satellites during three and $n$-body interactions
\citep[e.g.][]{1971Icar...15..186C,2006Natur.441..192A}. Planets may capture satellites
as they themselves interact in the presence of a planetesimal disk. Irregulars captured
earlier by gas drag and pull-down are supplanted by those captured during planet-planet
interactions. Such a scenario has been proposed for the origin of the Solar System's
irregular satellites within the context of the Nice model
\citep{2007AJ....133.1962N}. The strength of this model is that it results in similar
populations at each planet. In fact, in most model runs \citet{2007AJ....133.1962N} find
that Jupiter captures the least irregular satellites, because it undergoes the fewest
encounters with other planets. The prospects for lower mass planets harbouring swarms of
irregulars are therefore good.

Because we predict that swarms are most detectable around planets with large semi-major
axes, the probability of detection depends on the ability of planets to form on, or move
to such orbits. The recent discovery of planets around HR 8799
\citep{2008Sci...322.1348M} and Fomalhaut \citep{2008Sci...322.1345K} on such wide orbits
is encouraging, not only due to their very existence, but because both stars also harbour
planetesimal disks \citep{1984BAAS...16..483A,1986PASP...98..685S}. Planets that
originate on closer orbits must either scatter or migrate to such distances. If these
systems have planetesimal disks, satellite capture during the scattering and migration
process is likely. In such a scenario, the swarm's radial extent ($\eta$) cannot be much
larger than half the Hill radius at the planetary semi-major axis where the swarm was
captured (see also \S \ref{ss:fombdisc}).

The presence or absence of irregular swarms within the context of discovered planetary
systems should provide information about planet formation and evolution. For example, the
\citet{2007AJ....133.1962N} scenario requires the apparently specialised circumstances of
a set of giant planets that are destabilised and interact in the presence of a
planetesimal disk, but appears to be a robust way to capture irregulars. Such a scenario
suggests that swarms may be discovered around planets in multiple systems that harbour
debris disks (such as HR8799), especially those where scattering rather than migration
has occurred. On the other hand, if capture of irregulars is associated with major
planetesimal depletion events \citep[as proposed by][]{2007AJ....133.1962N}, systems with
irregular swarms may tend \emph{not} to have visible debris disks. If primordial
irregulars are formed in all planetary systems, the properties of swarm-harbouring
systems may be much more general. Therefore, it will be interesting to study the expected
detection outcomes for a range of capture and formation mechanisms.

As noted previously, there are effects beyond the scope of our model that can affect our
predictions. More theoretical effort should be made to understand which are important and
how they affect our conclusions, in particular the effect of radiation forces on
dust. Efforts should be made to detect dust clouds around Solar System giant planets, not
only for their intrinsic interest, but to provide an empirical conversion between the
mass in irregulars and the surface area in dust to calibrate our model predictions.

\subsection{Fomalhaut b}\label{ss:fomb}

An interesting application of our model is the planet orbiting the nearby star Fomalhaut,
which also harbours a narrow ring-like circumstellar debris disk
\citep{2008Sci...322.1345K}. The planet was predicted to exist based on the elliptical
shape of the debris ring \citep{2005Natur.435.1067K}. Using the sharpness of the inner
edge of the dust ring \citet{2006MNRAS.372L..14Q} estimated the planet's orbit and a
range of possible masses between Neptune's and Saturn's. In fact the structure of the
ring continues to provide the most stringent constraints on the planet's mass, with more
recent modelling providing an estimate of $<$3\,\mj\ if this planet is the ``sole
sculptor'' of the disk \citep{2009ApJ...693..734C}.

\citeauthor{2009ApJ...693..734C} note that Fomalhaut b need not be the only sculptor of
the belt. This point is in part made because the derived orbital velocity for Fomalhaut b
is marginally inconsistent with that expected if the disk and planet are apsidally
aligned. With Fomalhaut b only observed at two epochs, this inconsistency is a minor
issue at present, but highlights the possibility that another planet may be partly
responsible for the offset and truncation of the debris disk, and Fomalhaut b may be much
less massive.

Another reason the mass is so poorly constrained at present is the lack of information
about the planet's spectrum \citep{2008Sci...322.1345K}. Thus far it has only been
detected in two bands (0.6 and 0.8$\mu$m), the first of which shows a factor 3 change in
brightness between measurements at two epochs separated by 2 years. Neglecting the
potential temporal evolution, these observations and the non-detections in other
wavebands more closely resemble reflected starlight than thermal emission from a planet.

These observations lead \citeauthor{2008Sci...322.1345K} to suggest that the planet hosts
a circumplanetary ring akin to Saturn's, which would extend to at least 20 Jupiter radii
for an assumed albedo of 0.4 to recover the observed fluxes. If the emission is indeed
scattered light from dust then the planet mass could be much less than 3\,\mj\ 
\citep[see also][]{2004A&A...420.1153A}. In this section we suggest that the
observed spectrum could indeed be reflected starlight, scattered off the dusty debris
from a circumplanetary swarm of irregular satellites.

A convenient parameter space for this problem is the cloud semi-major axis in units of
Hill radii ($\eta$) vs. the planet mass ($M_{\rm pl}$). The observed quantities constrain
$\eta$, because the planet is not resolved and the disk size is limited by optical
depth. The planet mass will be constrained because a circumplanetary swarm that evolves
like our model will not survive for the age of Fomalhaut over all parts of this parameter
space. That is, for a swarm orbiting a $\sim$Jupiter-mass planet to be unresolved it
would have small $\eta$, where it would be rapidly depleted to undetectable levels.

\subsubsection{Observational Constraints}

The scattered light model in \citet{2008Sci...322.1345K} requires a total cross-sectional
area of dust:
\begin{equation}\label{eq:stot}
  \sigma_{\rm{tot}}= 5.8 \times 10^{-4} (0.08/Q)(1/\cos{i}) \, {\rm AU}^2 \, ,
\end{equation}
where we assume $Q = 0.08$ (see \S \ref{ss:obs}) and $\cos{i}$ is the factor that would
be required if the geometry was that of a flat ring inclined $i$ to our line-of-sight,
assumed to be $\sim$1 as the model of \S \ref{s:mod} is optically thin.

The light from Fomalhaut b looks unresolved so the dust must be confined to a region
smaller than the Hubble PSF FWHM of 0.53\,AU. This region has an area of 0.22AU$^2$ so
the geometrical optical depth of the dust could be as low as $2.6 \times 10^{-3}$. The
resolution constraint means that $\eta r_{\rm{H}} \lesssim 0.53$ so $\eta <
0.6/M_{\rm{pl}}^{1/3}$, where we have allowed the dust cloud to be a factor of two larger
in extent to account for the concentration of brightness closer to the planet seen in
Figure \ref{fig:saturnpic}. As noted in \S \ref{s:mod}, another constraint on $\eta$
comes from the stability of circumplanetary orbits, $\eta < 0.5$.

Assuming that the dust is uniformly projected on the sky across $\pi (\eta
r_{\rm{H}})^2$, its geometrical optical depth is $\tau = \sigma_{\rm{tot}} / ( \pi (\eta
r_{\rm{H}})^2 )$. Since $\tau < 1$ the dust must be located at $\eta > 0.015 / M_{\rm
  pl}^{1/3}$, which sets a lower limit on planet mass of $26 \times 10^{-6}M_\oplus$
(i.e., 6 times less massive than Ceres and 80 times less massive than Pluto).

\begin{figure}
  \begin{center}
    \vspace{-0.2in}
    \hspace{-0.35in} \psfig{figure=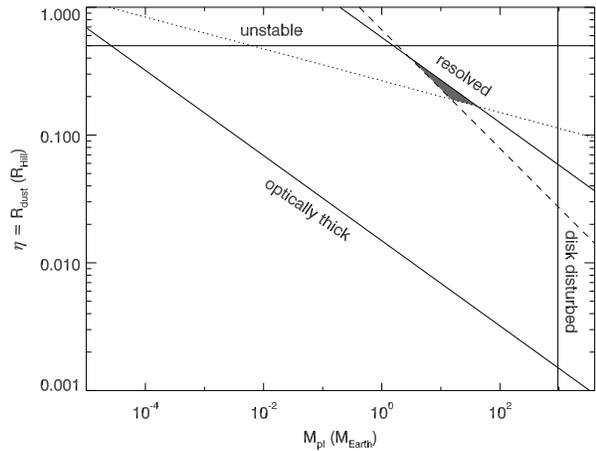,width=0.52\textwidth}
    \caption{Constraints on the location of the proposed irregular circumplanetary swarm
      (solid lines) and loci indicating lower limits to allowed regions with our model
      (dotted and dashed lines). The region where a satellite swarm could survive for
      200\,Myr is enclosed by the dashed and dotted lines, and the solid resolution line
      (shaded grey).}\label{fig:fomb}
  \end{center}
\end{figure}

The above constraints, and the dynamical constraint of $M_{\rm pl} \lesssim 3$\,\mj\ from
\citet{2009ApJ...693..734C}, are summarised by solid lines in Figure \ref{fig:fomb}. The
satellite cloud may reside anywhere in the region enclosed by the solid lines, which
spans over seven orders of magnitude in planet mass, and more than two in semi-major
axis. These constraints are independent of our model (except perhaps the 0.5\,\rh\
stability limit), and apply to any circumplanetary dust population. The $\sim$20\,\rj\
disk proposed by \citet{2008Sci...322.1345K} would lie in the lower right region of the
allowed space in Figure \ref{fig:fomb}.

Our aim is now to narrow this parameter space further. Because the collision rate depends
strongly on $\eta$, satellite clouds appear more likely to survive for the age of
Fomalhaut around planets much less massive than Jupiter.

Currently, the typically quoted age of Fomalhaut is 200\,Myr
\citep{1997ApJ...475..313B}. However, recent work suggests that the age may be closer to
400\,Myr (E. Mamajek, \emph{priv. comm.}). We adopt 200\,Myr as the age, and note below
the (small) difference an older age makes to our model.

\subsubsection{Collisional constraints}

To find where satellite clouds survive for the age of Fomalhaut, we set the collision
time ($1/R_{\rm cc}$) to 200\,Myr (i.e. it is collision limited). It remains to
substitute appropriate equations and estimates for parameters to reduce Equation
(\ref{eq:rccdc}) to $\eta$ as just a function of $M_{\rm pl}$. We use $M_\star =
2\,M_\odot$, $L_\star = 21\,L_\odot$, $\rho = 1500$\,kg m$^{-3}$, $a_{\rm pl} = 115$\,AU,
and $D_{\rm c} = 500$\,km. For comparison with Table \ref{t:ssdust}, the Hill
  radii for Earth, Neptune, and Jupiter-mass planets at this distance are 0.9, 2.3, and
  6.2AU respectively. We need to convert the observed \stot\ into \mtot, for which we
use equation (\ref{eq:mtot}). Because the effect of radiation pressure in setting \dmin\
varies over the parameter space, we also need equation (\ref{eq:dmin}), which gives the
mass as $M_{\rm tot} = 300 \, \sigma_{\rm tot} \eta^{0.35} / M_{\rm pl}^{0.23}$. The mass
in satellites is therefore about 0.06\,$M_\oplus$, or 5\,$M_{\rm moon}$ for $\eta = 0.2$
and $M_{\rm pl} = 10\,M_\oplus$, about 100-1000 times more massive than the initial
conditions in the \citet{2010AJ....139..994B} models, and similar to \S
\ref{ss:exo}. Such a low planet mass and higher stellar mass and luminosity result in
$D_{\rm min} \sim 300$\,\um, much higher than in the Solar System.

Substituting these parameters into equation (\ref{eq:rccdc}) yields a locus for a
collision limited satellite swarm around Fomalhaut b that reproduces the observed
\stot. This locus $\eta = 0.27 / M_{\rm pl}^{0.12}$ is shown in Figure \ref{fig:fomb} as
a dotted line. Because \dc\ may be smaller than our assumed 500\,km, but based on the
Solar System is unlikely to be significantly larger, the space above this line is also
allowed. Thus, the collision rate in concert with the resolution limit constrains the
planet to have a maximum mass of about 100\,$M_\oplus$. The minimum mass is lower than
the mass in satellites, and therefore physically unreasonable.

There is a further complication due to the large semi-major axis of Fomalhaut b, which
results in low collision velocities $v_{\rm rel} = 68 M_{\rm pl}^{1/3} / \sqrt{\eta}$ (in
m s$^{-1}$). A rough estimate of the largest object that can be destroyed by another of
the same size (and will therefore participate in the collisional evolution) can be
derived from equation (\ref{eq:qd}) and $X_{\rm c} = 1$ (\S \ref{ss:decay}), giving
$D_{\rm c} \approx v_{\rm rel}^{1.6}/28$ (in km). This estimate yields another collision
rate from equation (\ref{eq:rcc}) that applies when \dc\ is set by collision velocities
\begin{equation}
  R_{\rm cc} = 114 \, \frac{ M_{\rm tot} \, M_\star }
  { D_{\rm c}^{0.37} \, \rho \, ( \eta \, a_{\rm pl} )^3 \, M_{\rm pl} }
\end{equation}
in years$^{-1}$. This equation yields another locus in the $\eta$ vs. $M_{\rm pl}$
parameter space where the collision rate equals the system age of $\eta = 0.67 / M_{\rm
  pl}^{0.46}$. This limit is shown in Figure \ref{fig:fomb} as a dashed line. Though it
would appear that the dotted line allows swarms with $D_{\rm c} = 500$\,km to survive for
planet masses right down to $10^{-3}\,M_\oplus$, satellites this large cannot be
destroyed at the low collision velocities around such low mass planets. The dashed line
is a lower limit because \dc\ could be smaller than the smallest object that can be
destroyed. This limit therefore constrains the mass of Fomalhaut b to be more than a few
Earth masses.

These two loci combined with the previous constraints map out a region of parameter space
in Figure \ref{fig:fomb} where a satellite swarm could survive for 200\,Myr around
Fomalhaut b. The planet is $\sim$2--100 Earth masses and the swarm lies at 0.1-0.4 Hill
radii. The swarm mass is of order a few Lunar masses (but varies with planet mass due to
the changing \dmin, see above). The model predicts that a cloud of the observed
luminosity could survive around a more massive planet, but this possibility is excluded
because Fomalhaut b is unresolved. If Fomalhaut is 400\,Myr old, the allowed parameter
space is pushed even closer to the resolution limit, but does not change Figure
\ref{fig:fomb} significantly.

Fomalhaut b is marked in the lower panel of Figure \ref{fig:egpcoron} (based only on age
and orbit). As expected it lies within the region where swarms can be detected around a
20\,$M_\oplus$ planet with HST and JWST NIRCAM.

Of course there is considerable uncertainty in both our model and the parameters it uses,
but based on evidence that satellite swarms exist around Solar System planets, it at
least suggests that circumplanetary swarms should be considered a possibility around
extra solar planets. In the particular case of Fomalhaut b, the question of planet
vs. cloud can resolved if new observations show Fomalhaut b looks like a planetary
atmosphere. In the event that scattered light cannot be ruled out, both circumplanetary
rings \citep[e.g.][]{2008Sci...322.1345K} and swarms remain plausible
options.\footnote{Patient observers may someday find photometric phase variations that
  would be caused by an optically thick circumplanetary disk
  \citep{2004A&A...420.1153A}. However, other methods will likely become available before
  any variation can be found due to Fomalhaut b's $\sim$10$^3$ year period.}

\subsubsection{Discussion}\label{ss:fombdisc}

While our model provides a possible explanation of Fomalhaut b's apparently blue
spectrum, the provenance of such a configuration is unclear. Any scenario faces the
difficulty of explaining how Fomalhaut b came to be at such a large distance from
Fomalhaut itself. In a core accretion scenario, the planet presumably originates
somewhere much closer to the star \citep[e.g.][]{2008ApJ...673..502K}, and somehow
scatters or migrates to its current location. Because the Hill radius expands as the
planet moves outward, the satellite cloud would need to be captured while the planet was
orbiting at least $\sim$20\,AU from the star for it to reside at $\eta > 0.1$
now. Capture of satellites might happen as Fomalhaut b scatters off other planets or
migrates through a planetesimal disk on the way to its current location. The gas drag and
pull-down mechanisms are unlikely to operate because our predicted mass of Fomalhaut b is
insufficient for it to have a significant gaseous envelope.

Our predicted mass for Fomalhaut b is similar to or less than the mass of the main debris
ring itself, which is $\sim$3--300\,$M_\oplus$
\citep[e.g.][]{2002MNRAS.334..589W,2009ApJ...693..734C}. If the planet and ring masses
are similar, it is unlikely that Fomalhaut b is responsible for truncating and imposing
eccentricity on ring particles. Even if such a low-mass planet could reproduce the debris
ring structure, the relatively small chaotic zone width may require that the planet lie
closer to the ring than observed.

These issues do not necessarily pose a major problem for our
model. \citet{2009ApJ...693..734C} emphasise that Fomalhaut b may not actually be
responsible for sculpting Fomalhaut's debris ring, in part because the planet's orbit is
mildly inconsistent with the expected apsidal alignment of planet and ring particle
orbits. However, an as yet undiscovered object massive enough to sculpt the debris ring
may have more serious implications for the stability of Fomalhaut b.

\subsubsection{An alternative model}

The current poor constraint on Fomalhaut b's orbit allows for other interesting
possibilities. For example, its orbit may pass through the circumstellar ring. When the
planet is within the ring, planetesimal impacts will generate a surrounding cloud of
regolith, mantle, and planetesimal fragments. The material may free-fall back to the
planet as in the model of \citet{2002MNRAS.334..589W}, or in a picture more like the
formation of the Earth-Moon system or Kuiper belt binaries some fraction of the fragments
can remain in bound orbits. If Fomalhaut b spends a non-neglible fraction of its
  orbit within the ring, possible if the planet orbit and ring are coplanar, this
scenario provides a mechanism by which a population of objects bound to the planet may be
built up and replenished, thus avoiding the requirement that the swarm must survive for
the age of Fomalhaut to be observed. This way, even a small amount of mass launched into
orbit each time the planet passes through the ring will build up over time, eventually
reaching an equilibrium state where the bound mass is limited by collisions. Finding this
equilibrium mass is then simply a matter of estimating the rate at which mass is added to
the swarm and balancing it with the depletion rate. Because we expect the new satellites
to be launched into near-isotropic orbits (i.e. to look like an irregular satellite
swarm), the depletion rate is calculated from the model developed in \S \ref{s:mod}.

The rate at which new satellites are added to the swarm is uncertain, so we adopt the
approach taken by \citet{2002MNRAS.334..589W}; an impacting planetesimal can at most
launch its own mass of regolith and fragments to a significant distance from the
planet. For impacts by large planetesimals these collisions may be more akin to the
``graze and capture'' like scenario suggested for Pluto-Charon and Haumea's satellites
\citep[e.g.][]{2005Sci...307..546C,2010ApJ...714.1789L}. Not all planetesimals accreted
by the planet will launch their own mass in regolith and fragments away from the planet,
nor will all launched material end up in orbit. We therefore expect that the accretion
rate must be at least a few orders of magnitude higher than the loss rate for the
existence of a cloud to be feasible.

The gravitationally focused mass accretion rate can be calculated from
\begin{equation}\label{eq:macc}
  \dot{M}_{\rm acc} =
  \frac{M_{\rm ring}}{2\,\pi\,r_{\rm ring}^2\,dr_{\rm ring}\,i_{\rm ring}} \,
  \pi R_{\rm pl}^2 \, \left( 1 + \frac{v_{\rm esc}^2}{v_{\rm rel}^2} \right) \, v_{\rm rel}
\end{equation}
We assume $r_{\rm ring} = 141$\,AU, $dr_{\rm ring} = 25$\,AU, and $i_{\rm ring} =
0.026$\,rad \citep[using ring parameters from][]{2005Natur.435.1067K}. We assume the
largest ring planetesimal is 500km in diameter, which corresponds to a ring mass of about
75$M_\oplus$ \citep{2005Natur.435.1067K}, which is intermediate to the two cases
considered in \citet{2002MNRAS.334..589W}. The accretion rate of a 10\,$M_\oplus$ planet
with mass density 5000\,kg m$^{-3}$ is about $10^{-10} M_\oplus/$year. Because the planet
cannot spend all its time within the ring, we (arbitrarily) decrease the accretion rate
by a factor of two. The accretion rate would be much lower for a non-coplanar
  orbit, on which the planet would spend much less time (if any) within the ring. For a
10\,$M_\oplus$ planet with mass density 5000\,kg/m$^3$, and $v_{\rm rel} = 0.1\,v_{\rm
  K}$, Equation (\ref{eq:macc}) reduces to $2 \times 10^{-10} M_{\rm
  pl}^{4/3}\,M_\oplus/$yr.

The mass loss rate is simply the current mass times the collision rate, which for a swarm
in collisional equilibrium is also the inverse of the age (equation \ref{eq:dmdt}). More
generally, using Equations (\ref{eq:qd}), (\ref{eq:mtot}), and (\ref{eq:rccdc}), the mass
loss rate is
\begin{equation}\label{dmdtstot}
  \dot{M}_{\rm loss} =  0.002 \,
  \left[ \frac{\rho^{0.37} \, D_{\rm min}^{1.4} \, M_\star^{1.38} \, f_{v_{\rm rel}}^{2.27}}
    {M_{\rm pl}^{0.24} \, \left(a_{\rm pl} \, \eta\right)^{4.13}} \right] \,
  \sigma_{\rm tot}^2
\end{equation}
in $M_\oplus /$ year. For our purposes here, the mass loss rate is simply proportional to
the square of the observed $\sigma_{\rm tot}$. That is, more mass is lost and at a faster
rate for higher \stot\ or \mtot. All parameters in the large parenthesis depend on the
particular model assumptions. The rate is independent of the maximum planetesimal size
\dc. This independence can be viewed as due to our assumption of a self-similar size
distribution, where the mass lost is the same in each logarithmic size bin. If the small
end of the size distribution is fixed, larger \dc\ means more mass is available to be
depleted (\mtot), but collisions are less frequent and the largest objects are stronger.

Evaluating Equation (\ref{dmdtstot}) for Fomalhaut b yields $\dot{M}_{\rm loss} = 5.5
\times 10^{-12} / \left( \eta^{3.4} \, M_{\rm pl}^{0.7} \right)$. Using the system age
and a planet with a collision limited swarm of mass $0.06\,M_\oplus$ from the previous
section yields $3 \times 10^{-10}\,M_\oplus/$year, in agreement with the general
expression for these parameters.

\begin{figure}
  \begin{center}
    \vspace{-0.2in}
    \hspace{-0.35in} \psfig{figure=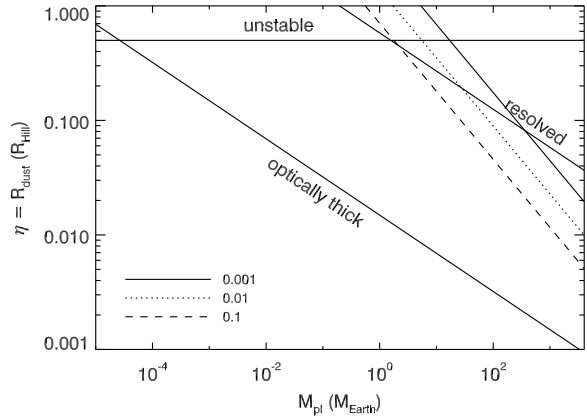,width=0.52\textwidth}
    \caption{Lines in $\eta$ vs. \mpl\ parameter space where swarms can be replenished by
      accretion with different efficiency parameters $f_{\rm acc}$ (shown in
      legend). Though the ``disk disturbed'' limit is not drawn, we expect it to lie at a
      much lower planet mass than in Figure \ref{fig:fomb}.}\label{fig:cre}
  \end{center}
\end{figure}

Using these two expressions for the mass accretion and loss rates and assuming an
efficiency parameter $f_{\rm acc}$, a swarm will replenished as fast as it decays when
$f_{\rm acc} \, \dot{M}_{\rm acc} = \dot{M}_{\rm loss}$. Figure \ref{fig:cre} shows where
this equation is satisfied for different $f_{\rm acc}$. The line where the planet mass is
too large to be consistent with the observed debris ring structure is omitted because a
planet that passes through ring will perturb it in a different way. This line would move
to lower planet masses, and constrain this model. Another constraint on the upper planet
mass would come from the mass above which a significant atmosphere is present, meaning
that ring planetesimals are engulfed, rather than launch regolith and fragments. However,
an atmosphere under periodic bombardment would be more prone to thermal escape due to
heating.

The location of the lines in Figure \ref{fig:cre} suggest that replenishing a swarm from
ring planetesimals is difficult. Even if a tenth of the accreted mass is launched into
orbit the planet must be more than an Earth mass, which seems likely to significantly
perturb the main ring on a timescale less than 200Myr. One way to lower the planet mass
but maintain the accretion rate is if the planet is a binary. The cross section for an
interaction with a passing planetesimal scales with the binary semi-major axis and is
therefore much larger than the physical size of either object
\citep{1975AJ.....80..809H,1989AJ.....98..217L}. Such an interaction does not guarantee a
collision, but allows a greater chance of one for the same (total) planet mass. This may
allow a binary to launch sufficient material into a swarm, yet also have a low enough
mass to minimise dynamical perturbations to Fomalhaut's debris ring.

\section{Summary and Conclusions}\label{s:conc}

The purpose of this paper is to derive a simple but general model for collisional
evolution of irregular satellites and apply it to planets in the Solar System and
elsewhere. Our model uses the particle-in-a-box formalism to describe the collision rate
of the largest objects in a satellite swarm. A model size distribution allows the mass
evolution to be converted to the surface area in dust and flux densities. Though the
model can be developed further in many ways, a comparison with the Solar System irregular
satellites shows that our model is reasonable.

Application of our model to the Solar System suggests that the bulk of grains may be lost
to the planet or interplanetary space when the cloud is collision dominated. It may be
that deposition on regular satellites only becomes important at later times when PR drag
starts to dominate grain orbital evolution. Some level of dust must be present if the
irregular satellites are still grinding down, which we suggest may be at detectable
levels at any of the Solar System's outer planets.

A remarkable feature of irregular satellite swarm evolution is its relative insensitivity
to planet mass. Swarms are nearly as bright and last nearly as long around Neptune-mass
planets as around Jupiter-mass planets. In contrast to extrasolar planets, we find that
satellite swarms are most visible at optical wavelengths around planets that orbit at
many tens of AU from their parent stars. Lower mass planets without gaseous envelopes
cannot capture swarms by gas drag and pull-down, but dynamical mechanisms can still
operate. There is an optimum distance for detection, which arises because swarms around
planets on close orbits decay too rapidly and swarms around planets on distant orbits
take too long to collide.

We propose a plausible model for a satellite swarm around Fomalhaut b. This model
provides an alternative explanation for the spectrum of the planet being consistent with
reflected starlight. In this picture Fomalhaut b is predicted to be $\sim$1--100 Earth
masses, further illustrating that planet mass is relatively unimportant for the evolution
of irregular satellite swarms. The allowed parameter space for the model lies very close
to the resolution limit, so observations at higher resolution (${\rm FWHM} < 0.53$\,AU or
69\,mas) should test our model. However, the order-of-magnitude spirit of our model and
large uncertainties in model parameters that affect our allowed parameter space, such as
satellite size, composition and strength, mean that the best way to test our model in
this particular instance is to ascertain whether Fomalhaut b has the spectrum expected
from a planet.

We briefly outline another possibility for a swarm around Fomalhaut b based on
speculation that it could pass through Fomalhaut's large debris ring. In this case, a
swarm of objects is maintained by regolith and fragments launched by planetesimal
impacts. The $\sim$Earth mass planet required by such a model may be too high to avoid
large (and unobserved) perturbations to the main ring. This issue may be somewhat
alleviated if the planet is actually a binary, which would enhance the accretion rate and
allow a lower total planet mass.

The prospects for detecting dust created by irregular satellite collisions, both in the
Solar System and around planets orbiting other stars appear good. Along with further
characterisation of Fomalhaut b, the four populations of irregulars on our doorstep seem
a good place to start.

\section*{Acknowledgments}
We are grateful to the Isaac Newton Institute for Mathematical Sciences in Cambridge
where part of this work was carried out during the Dynamics of Discs and Planets research
programme. We would also like to thank Paul Kalas for helpful comments and sharing
preliminary observations of Fomalhaut b, Eric Mamajek for sharing his results on
Fomalhaut's age, and the referee Bill Bottke for a constructive review.

\bibliography{ref} \bibliographystyle{astroads}

\end{document}